\documentclass[10pt]{article}
\usepackage{fullpage}
\usepackage{amsmath}
\usepackage{amsfonts}
\usepackage{amssymb}
\usepackage[dvips]{epsfig}
\usepackage{color}

\def\##1{\underline{#1}}
\def\=#1{\underline{\underline{#1}}}

\def\+#1{\underline{\bf #1}}
\def\*#1{\underline{\underline{\bf #1}}}

\def\~#1{{\tilde #1}}

\def\.{\mbox{ \tiny{$^\bullet$} }}

\def\eps{\epsilon}

\def\c#1{\cite{#1}}
\def\l#1{\label{#1}}
\def\r#1{(\ref{#1})}

\def\le{\left(}
\def\ri{\right)}
\def\les{\left[}
\def\ris{\right]}

\renewcommand{\thefootnote}{\fnsymbol{footnote}}

\begin{document}

\begin{center}

\LARGE{ {\bf Giant dielectric anisotropy via homogenization}}
\end{center}
\begin{center}
\vspace{5mm} \large

 Tom G. Mackay\footnote{E--mail: T.Mackay@ed.ac.uk}\\
{\em School of Mathematics and
   Maxwell Institute for Mathematical Sciences\\
University of Edinburgh, Edinburgh EH9 3JZ, UK}\\
and\\
 {\em NanoMM~---~Nanoengineered Metamaterials Group\\ Department of Engineering Science and Mechanics\\
Pennsylvania State University, University Park, PA 16802--6812,
USA}
\normalsize

\end{center}

\bigskip
\renewcommand{\thefootnote}{\arabic{footnote}}
\setcounter{footnote}{0}

\bigskip

\vspace{5mm}

\begin{center}
\vspace{1mm} {\bf Abstract}

\end{center}

\vspace{4mm}

A random mixture of two isotropic dielectric materials, one composed of oriented spheroidal particles of relative permittivity $\eps_a$ and the other composed of oriented spheroidal particles of relative permittivity $\eps_b$, was considered in the  long wavelength regime. The permittivity dyadic of the resulting homogenized composite material (HCM) was estimated using the Bruggeman homogenization formalism. The HCM was an orthorhombic biaxial material if the symmetry axes of the two populations of spheroids were mutually perpendicular and a uniaxial material if these two axes were mutually aligned.
The degree of anisotropy of the HCM, as gauged by the ratio of the eigenvalues of the HCM's permittivity dyadic, increased as the shape of the constituent particles became more eccentric.  The greatest degrees of HCM anisotropy were achieved for the limiting cases wherein the constituent particles were shaped as needles or discs. In these instances explicit formulas for the HCM anisotropy were derived from the dyadic Bruggeman equation. Using these formulas it was found that the degrees of HCM anisotropy are proportional to $\sqrt{\eps_b}$ or $\eps_b$, at fixed values of volume fraction and $\eps_a$, for $\eps_b > \eps_a$. Thus, in principle, there is no limit to degree of anisotropy that may be attained via homogenization. In practice, the degree of anisotropy would be limited by the available value of $\eps_b$ (and/or $\eps_a$).

\vspace{5mm}

\noindent {\bf Keywords:} Bruggeman homogenization formalism; needle--shaped particles; disc--shaped particles; giant anisotropy

\vspace{5mm}

%\newpage

\section{Introduction}

While nature provides us with a great many anisotropic materials \c{Highly_anisotropic_crystals,Moriaki}, there are occasions when a material with a specific degree of anisotropy may be required  but none is readily available. On such occasions we may turn to engineered composite materials. Rather exotic dielectric anisotropies have been central to  recent developments involving
 nanostructured composite materials  which support  electromagnetic phenomenons such as negative refraction \c{Cui}, optical cloaking \c{Cloak},  null reflection \c{Zero_reflection}, and omnidirectional radiation \c{Cui_PRL}, for examples. Most significantly, the incorporation of dielectric anisotropy can enable  non--magnetic composite materials to support negative refraction  \c{Narimanov,Mortensen}.
Another notable area where dielectric anisotropy plays a key role is in the development of material analogues for the electromagnetic properties of certain curved spacetime scenarios, such as rotating black holes \c{Smolyaninov_NJP,Lu_JAP}, Schwarzschild-(anti-)de Sitter spacetime
\c{ML_PRB}, and cosmic spinning strings \c{Spinning_string}, as well as material analogues of quantum electrodynamic vacuum \c{QED}. In particular, high degrees of dielectric anisotropy are needed to represent regions of large spacetime curvature, close to singularities or event horizons, for examples.

Biaxial or uniaxial
anisotropy can be attained by homogenizing  composite materials which are composed of  oriented  constituent particles characterized by certain symmetries, such as cylindrical \c{Mortensen} or ellipsoidal \c{Homog_ellipsoids}   symmetry.
In the following, we investigate a means  of achieving very high degrees of dielectric anisotropy, in a controllable manner, through the homogenization of remarkably simple component
materials, namely isotropic dielectric materials composed of spheroidal particles. The approach taken is based on the well--established Bruggeman homogenization formalism \c{Ward,Goncharenko}.

A note concerning notation: In the following, 3--vectors are single underlined while 3$\times$3 dyadics are double underlined.
Unit vectors are signified by the $\hat{}$ symbol. Thus, unit vectors aligned with the coordinate axes are written as
$\hat{\#x}$,  $\hat{\#y}$ and $\hat{\#z}$. The identity and null 3$\times$3 dyadics are denoted by $\=I$ and $\=0$, respectively.

\section{Homogenization preliminaries} \l{homog_sec}

\subsection{Component materials} \l{component_materials}

Let us study the homogenization of two isotropic dielectric component materials, namely material $a$ which is characterized by the relative permittivity $\eps_a$ and  material $b$ which is characterized by the relative permittivity $\eps_b$. It assumed that
 the component materials are lossless and that $\eps_{a,b} > 0$.\footnote{Homogenization formalisms can yield results which are not physically plausible in the $\eps_a \eps_b < 0$ regime \c{M_Ag,Milton}.}
 Both component materials are particulate in nature; their constituent particles are taken to be spheroidal in shape (and limiting cases of these spheroidal shapes are also considered). The component materials are randomly mixed together to form a composite material, with component material $a$ having the volume fraction $f_a$ and material $b$ the volume fraction $f_b = 1- f_a$.
In the composite material, all material $a$ spheroidal particles are assumed to have the same shape and orientation, and likewise all material $b$ spheroidal particles are assumed to have the same shape and orientation. The surface of each constituent spheroidal particle of type $a$ or type $b$, relative to its centroid,
is traced out by the
 vector
\begin{equation}
\#r_{\,s} (\theta, \phi) = \eta \, \=U_{\,\ell} \. \hat{\#r} (\theta, \phi), \qquad \quad \le \ell = a, b \ri.
\end{equation}
Here $ \hat{\#r} $ is the radial unit vector with its origin coinciding with the
spheroid's centroid; it is specified in terms of the spherical polar coordinates
$\theta$ and $\phi$. The  dyadic
\begin{equation} \l{Ushape}
 \=U_{\, \ell} =  \frac{1}{\sqrt{\gamma_\ell}} \le \=I - \hat{\#c}_{\, \ell} \, \hat{\#c}_{\, \ell} \ri + \gamma_\ell \, \hat{\#c}_{\, \ell} \, \hat{\#c}_{\, \ell}\,,
 \qquad \quad \le \ell = a, b \ri,
\end{equation}
characterizes the
spheroidal shape and orientation; herein
  the unit vector
$\hat{\#c}_{\, \ell}$ is aligned with the spheroid's axis of rotational
symmetry. The eccentricity of the spheroid is captured by the positive-valued parameter $\gamma_\ell$;
for the degenerate case $\gamma_\ell = 1$ the spheroid takes the form of a sphere.
 %(and the limiting cases $\gamma_\ell \to 0$ and
%$\gamma_\ell \to \infty$ are taken up in \S\ref{limiting_cases}).
The linear dimensions of the spheroid are
fixed by the positive--valued size parameter $\eta$.

In the following sections two cases are investigated: (i) the case where the component material $a$ spheroids are aligned perpendicularly to the component material $b$ spheroids  (i.e., $ \hat{\#c}_{\, a} \. \hat{\#c}_{\, b} = 0$); and (ii)
  the case where the component material $a$ spheroids and the component material $b$ spheroids have the same alignment (i.e., $ \hat{\#c}_{\, a} \. \hat{\#c}_{\, b} = 1$).
  To be specific, let us choose $\hat{\#c}_{\, a} = \hat{\#z}$ and $\hat{\#c}_{\, b} = \hat{\#y}$ for case (i), and
$\hat{\#c}_{\, a} = \hat{\#c}_{\, b} = \hat{\#z}$ for case (ii).
For simplicity,  the eccentricity of the component material $a$ spheroids is taken to be the same as that for the component material $b$ spheroids; accordingly, we introduce the eccentricity parameter $\gamma \equiv \gamma_a = \gamma_b$. Schematic representations of these cases (i) and (ii) are provided in Fig.~\ref{fig1}.

\subsection{Homogenized composite material}

The composite material described in \S\ref{component_materials} may be regarded as being effectively homogeneous provided that the constituent spheroidal particles are much smaller than  the wavelength(s) under consideration.
Unlike the component materials, the corresponding homogenized composite material (HCM) is an anisotropic dielectric material \c{EAB}. The anisotropic nature of the HCM stems from the geometry of its  oriented spheroidal constituent particles.
For both cases (i) and (ii),
 the relative permittivity dyadic of the HCM may be represented by the general form
\begin{equation} \l{eHCM}
 \=\eps_{\, HCM} =  \eps_x \, \hat{\#x} \, \hat{\#x} + \eps_y \, \hat{\#y} \, \hat{\#y} + \eps_z \, \hat{\#z} \, \hat{\#z}\,.
\end{equation}
 For case (i) there are three distinct relative permittivity parameters, namely $\eps_x$, $\eps_y$ and $\eps_z$, and the corresponding  HCM is an orthorhombic biaxial material.
 For case (ii) there are only two distinct relative permittivity parameters, namely $\eps_x$ and $\eps_z$, since here $\eps_x = \eps_y$;  and the corresponding HCM is a uniaxial dielectric material.

 The  relative permittivity dyadic of the HCM
may be estimated by means of the widely--used Bruggeman homogenization formalism \c{Ward,Goncharenko}.
This process involves extracting $ \=\eps_{\, HCM}$ from the dyadic Bruggeman equation \c{EAB}
\begin{equation}
f_a \, \=\alpha_{\,a} + f_b \, \=\alpha_{\,b}  = \=0\,, \l{Br}
\end{equation}
which is expressed in terms of the polarizability density dyadics
\begin{equation}
\=\alpha_{\,\ell } = \le \eps_\ell \=I - \=\eps_{\,HCM} \ri \.\les \, \=I
+
 \=D_{\,\ell} \. \le \eps_\ell \=I - \=\eps_{\,HCM} \ri \,\ris^{-1}, \qquad
(\ell = a,b). \l{polar}
\end{equation}
The depolarization dyadics $\=D_{\,\ell}$ herein are given by
the double integrals
 \c{M97,MW97}
  \begin{equation} \l{D_dint}
\=D_{\, \ell} = \frac{1}{4 \pi} \int^{2 \pi}_\phi d\phi \int^\pi_\theta d \theta \, \sin \theta\, \frac{ \le \=U^{-1}_{\,\ell} \. \hat{\#r} \ri \le \=U^{-1}_{\,\ell} \. \hat{\#r} \ri}{\le \=U^{-1}_{\,\ell} \. \hat{\#r} \ri \. \=\eps_{\,HCM} \. \le \=U^{-1}_{\,\ell} \. \hat{\#r} \ri} ,\qquad \qquad \le \ell = a, b \ri.
  \end{equation}
The components of $\=D^{}_{\, \ell}$ may be  expressed in terms of incomplete elliptic functions
when the HCM is an orthorhombic biaxial material (i.e., case (i)) \c{W98}, and in terms of inverse hyperbolic and trigonometric functions
when the HCM is a uniaxial material (i.e., case (ii)) \c{M97}; further details are provided in the Appendix.

Due to
the nonlinearity of the dyadic Bruggeman equation \r{Br}, numerical techniques are usually needed to deliver $ \=\eps_{\, HCM}$
when the HCM is an anisotropic (or bianisotropic)  material \c{PiO}. However, 
as presented in \S\ref{limiting_cases},
for certain limiting  cases explicit solutions can be derived.

\section{Spheroidal constituent particles} \l{spheroidal_sec}

By means of some representative numerical examples, let us explore the anisotropy that may be induced through homogenizing the assembly of oriented spheroidal particles described in \S\ref{component_materials}. We focus on the quantities $\eps_x/ \eps_y$, $\eps_y/ \eps_z$ and
$\eps_x/ \eps_z$ which provide measures of the degrees of anisotropy exhibited by the HCM.

Suppose that $\eps_a = 1.5$ and $\eps_b = 12$. For case (i), wherein
$\hat{\#c}_{\, a} = \hat{\#z}$ and $\hat{\#c}_{\, b} = \hat{\#y}$, the quantities $\eps_x/ \eps_y$, $\eps_y/ \eps_z$ and
$\eps_x/ \eps_z$ are plotted versus the eccentricity parameter $\gamma \in\le 0.1, 6 \ri$ in Fig.~\ref{fig2}. Here the
volume fraction $f_a = 0.7$ (green, dashed curves), 0.4 (red, solid curves) and 0.1 (blue, broken dashed curves).
The magnitudes of $\eps_x/ \eps_y$, $\eps_y/ \eps_z$ and
$\eps_x/ \eps_z$  diverge from unity as $\gamma$ diverges from unity.
While the HCM is clearly biaxial, for $f_a = 0.1$ the quantity $\eps_x/\eps_y$ is approximately equal to one at all values of $\gamma$ considered; this indicates that when the concentration of component material $a$ is very small,
the electromagnetic properties of the HCM are very much dominated by component material $b$ and accordingly the HCM is nearly uniaxial.
At $\gamma = 1 $, we have $\eps_x = \eps_y = \eps_z$ and the HCM is isotropic.

We repeat the calculations of Fig.~\ref{fig2} for case (ii), wherein $\hat{\#c}_{\, a} = \hat{\#c}_{\, b} = \hat{\#z}$.
The anisotropy parameter $\eps_y / \eps_z$  ($ \equiv \eps_x/\eps_z$) is plotted versus $\gamma$ in Fig.~\ref{fig3}. The trends in Fig.~\ref{fig3}
are similar to those in Fig.~\ref{fig2}, but for Fig.~\ref{fig3} the maximum   values of $\eps_y/\eps_z$  are larger than the maximum values  of  $\eps_x/ \eps_y$, $\eps_y/ \eps_z$ and
$\eps_x/ \eps_z$ for Fig.~\ref{fig2}; and likewise the minimum   values of $\eps_y/\eps_z$  are smaller than the  minimum values  of  $\eps_x/ \eps_y$, $\eps_y/ \eps_z$ and
$\eps_x/ \eps_z$ for Fig.~\ref{fig2}. Thus, we deduce that greater degrees of anisotropy can be achieved when the alignments of the two types of constituent spheroids are the same as compared to the corresponding scenario when the alignments of the two types of constituent spheroids are mutually perpendicular.

For both cases (i) and (ii),  the HCM becomes more anisotropic as the constituent particles become more eccentric in shape. However, it would appear  from Figs.~\ref{fig2} and \ref{fig3} that there are limits upon the degrees of anisotropy that can be achieved through varying the eccentricity parameter $\gamma$ and that these limits depend upon the volume fractions of the component materials. We pursue this matter in \S\ref{limiting_cases}.

\section{Limits to anisotropy} \l{limiting_cases}

What is the greatest degree of anisotropy than can be achieved by homogenizing an assembly of oriented spheroidal particles? In order to address this question, the limits $\gamma \to \infty$ and $\gamma \to 0$ are considered  in the following \S\ref{needles_sec} and \S\ref{disc_sec}, respectively. The depolarization dyadics degenerate to simple forms in these limits, as has been demonstrated in earlier works by a direct analysis of  Eqs.~\r{D_dint} \c{WM_Needles_Pillboxes} or by considering the corresponding eigenfunction expansion cast in cylindrical coordinates \c{Cottis}. These simplified forms for the depolarization dyadics render the dyadic Bruggeman equation \r{Br} amenable to analysis. % (unlike the situation in \S\r{spheroidal_sec}).

\subsection{Needle--shaped constituent particles} \l{needles_sec}

In the limit $\gamma \to \infty$ the constituent particles may be regarded as needle--shaped. For case (i) wherein
$\hat{\#c}_{\, a} = \hat{\#z}$ and $\hat{\#c}_{\, b} = \hat{\#y}$, the depolarization dyadics \r{D_dint} reduce to  \c{WM_Needles_Pillboxes}
\begin{equation}
\left.
\begin{array}{l}
\=D_{\,a} = \displaystyle{\frac{1}{\eps_x - \eps_y} \les \le 1- \sqrt{\frac{\eps_y}{\eps_x}} \ri \hat{\#x} \, \hat{\#x} +
 \le \sqrt{\frac{\eps_x}{\eps_y}} -1 \ri \hat{\#y} \, \hat{\#y} \ris }
 \vspace{8pt} \\
\=D_{\,b} = \displaystyle{\frac{1}{\eps_z - \eps_x} \les
 \le \sqrt{\frac{\eps_z}{\eps_x}} -1 \ri \hat{\#x} \, \hat{\#x} + \le 1- \sqrt{\frac{\eps_x}{\eps_z}} \ri \hat{\#z} \, \hat{\#z} \ris }
 \end{array}
 \right\}.
\end{equation}
We investigate numerically the corresponding estimates provided by the Bruggeman homogenization formalism. As for Figs.~\ref{fig2} and \ref{fig3}, let us fix $\eps_a = 1.5$. The quantities
$\eps_x/ \eps_y$, $\eps_y/ \eps_z$ and
$\eps_x/ \eps_z$ are plotted versus the relative permittivity $\eps_b \in\le 0.01, 200 \ri$ in Fig.~\ref{fig4}. As previously, the
volume fraction $f_a = 0.7$ (green, dashed curves), 0.4 (red, solid curves) and 0.1 (blue, broken dashed curves). The magnitudes of $\eps_x/ \eps_y$, $\eps_y/ \eps_z$ and
$\eps_x/ \eps_z$ diverge from unity as $\eps_b$ diverges from 1.5 (the value of $\eps_a$). In the limit $\eps_b \to 1.5$ the HCM becomes an isotropic dielectric material, regardless of the volume fraction. For the range of $\eps_b$ and $f_a$ values considered in Fig.~\ref{fig4}, the magnitudes of $\eps_x/ \eps_y$, $\eps_y/ \eps_z$ and
$\eps_x/ \eps_z$ lie within the interval $\le 0.15, 4.4\ri$.

Turning to case (ii) wherein
$\hat{\#c}_{\, a} = \hat{\#c}_{\, b} = \hat{\#z}$, the depolarization dyadics \r{D_dint} reduce to
\begin{equation}
\=D_{\,\ell} = \frac{1}{2 \eps_x }  \le  \hat{\#x} \, \hat{\#x} +
  \hat{\#y} \, \hat{\#y} \ri, \qquad \quad \le \ell = a, b \ri.
\end{equation}
These particularly simple forms for $\=D_{\, a, b}$ allow an explicit solution to be extracted from the dyadic Bruggeman equation
\r{Br}. Thus, we find
\begin{equation} \l{eps_na}
\left.
\begin{array}{l}
\eps_x = \eps_y = \displaystyle{\frac{ \le f_b-f_a \ri \le \eps_b - \eps_a \ri + \sqrt{\les \le f_b-f_a \ri \le \eps_b - \eps_a \ri \ris^2 + 4 \eps_a \eps_b} }{2}} \vspace{8pt} \\
\eps_z = f_a \eps_a + f_b \eps_b
\end{array}
\right\}.
\end{equation}
Let us illustrate the anisotropic nature of the HCM by repeating the calculations of Fig.~\ref{fig4} but with
$\hat{\#c}_{\, a} = \hat{\#c}_{\, b} = \hat{\#z}$. The corresponding plots for the quantity $\eps_y / \eps_z$ are presented in Fig.~\ref{fig5}. We see that the magnitude of $\eps_y / \eps_z$  decreases steadily from unity as $\eps_b$ diverges from 1.5 (the value of $\eps_a$). Thus, the HCM becomes increasingly anisotropic as $\eps_b$  deviates from $\eps_a$.
For the range of $\eps_b$ and $f_a$ values considered in Fig.~\ref{fig5}, the magnitude of  $\eps_y/ \eps_z$  lies within the interval $\le 0.05, 1\ri$.

Further insight into this matter may be gained by considering the expressions for $\eps_y$ and $\eps_z$ given in Eqs.~\r{eps_na}. Combining these expressions for the instance  $f_a = f_b = 0.5$, we find
\begin{equation}
\frac{\eps_z}{\eps_y} = \frac{1}{2} \le \sqrt{\frac{\eps_a}{\eps_b}} + \sqrt{\frac{\eps_b}{\eps_a}} \ri \to \infty \quad \mbox{as} \quad \eps_b \to \left\{ \begin{array}{l} 0 \\ \infty \end{array} \right. \quad \mbox{for fixed $\eps_a$.}
\end{equation}
That is, the degree of anisotropy, as gauged by $\eps_z/\eps_y$, can increase without limit as $\eps_b$ increasingly deviates from $\eps_a$, and  the degree of anisotropy is proportional to $\sqrt{\eps_b}$ for $\eps_b > \eps_a$ and proportional to $1/\sqrt{\eps_b}$ for $\eps_b < \eps_a$.

\subsection{Disc--shaped constituent particles} \l{disc_sec}

In the limit $\gamma \to 0$ the constituent particles may be regarded as disc--shaped.  The corresponding depolarization dyadics \r{D_dint}, for case (i) wherein
$\hat{\#c}_{\, a} = \hat{\#z}$ and $\hat{\#c}_{\, b} = \hat{\#y}$,
 reduce to  \c{WM_Needles_Pillboxes}
\begin{equation}
\left.
\begin{array}{l}
\=D_{\,a} = \displaystyle{\frac{1}{\eps_z} \hat{\#z} \, \hat{\#z}}
 \vspace{8pt} \\
\=D_{\,b} = \displaystyle{\frac{1}{\eps_y} \hat{\#y} \, \hat{\#y}}
 \end{array}
 \right\}.
\end{equation}
For these  simple depolarization dyadic forms, the following  explicit solution can be extracted from the dyadic Bruggeman equation
\r{Br}:
\begin{equation} \l{eps_pp}
\left.
\begin{array}{l}
\eps_x = f_a \eps_a + f_b \eps_b
\vspace{8pt} \\
\eps_y =  \displaystyle{\frac{ \le f_b-f_a \ri  \eps_b  + \sqrt{ \le f_b-f_a \ri^2  \eps^2_b  + 4 f_a f_b \eps_a \eps_b} }{2 f_b}} \vspace{8pt} \\
\eps_z =  \displaystyle{\frac{ \le f_b-f_a \ri  \eps_a  + \sqrt{ \le f_b-f_a \ri^2  \eps^2_a  + 4 f_a f_b \eps_a \eps_b} }{2 f_a}}
\end{array}
\right\}.
\end{equation}
Let us illustrate this solution numerically.
  As for Figs.~\ref{fig2}--\ref{fig5}, we set $\eps_a = 1.5$. The quantities
$\eps_x/ \eps_y$, $\eps_y/ \eps_z$ and
$\eps_x/ \eps_z$ are plotted versus the relative permittivity $\eps_b \in\le 0.01, 200 \ri$ in Fig.~\ref{fig6}. As previously, the
volume fraction $f_a = 0.7$ (green, dashed curves), 0.4 (red, solid curves) and 0.1 (blue, broken dashed curves).
The general trends are similar to those exhibited in Figs.~\ref{fig4} for needle--shaped particles; that is,
the magnitudes of $\eps_x/ \eps_y$, $\eps_y/ \eps_z$ and
$\eps_x/ \eps_z$ diverge from unity as $\eps_b$ diverges from 1.5 (the value of $\eps_a$),  regardless of the volume fraction, with the HCM becoming an isotropic dielectric material  in the limit $\eps_b \to 1.5$. For the range of $\eps_b$ and $f_a$ values considered in Fig.~\ref{fig6}, the magnitudes of $\eps_x/ \eps_y$, $\eps_y/ \eps_z$ and
$\eps_x/ \eps_z$ lie within the interval $\le 0.2, 4.4\ri$.

We can better appreciate the anisotropic nature of the HCM here
 by considering the  expressions for $\eps_x$, $\eps_y$ and $\eps_z$ given in Eqs.~\r{eps_pp} for the instance  $f_a = f_b = 0.5$. We find that
\begin{equation}
\frac{\eps_x}{\eps_y} = \frac{\eps_x}{\eps_z} = \frac{1}{2} \le \sqrt{\frac{\eps_a}{\eps_b}} + \sqrt{\frac{\eps_b}{\eps_a}} \ri \to \infty \quad \mbox{as} \quad \eps_b \to \left\{ \begin{array}{l} 0 \\ \infty \end{array} \right. \quad \mbox{for fixed $\eps_a$,}
\end{equation}
and $\le \eps_y / \eps_z \ri = 1$.
That is, the degree of anisotropy, as gauged by $\eps_x/\eps_y$ and $\eps_x / \eps_z$, can increase without limit as $\eps_b$ increasingly deviates from $\eps_a$, and  the degree of anisotropy is proportional to $\sqrt{\eps_b}$ for $\eps_b > \eps_a$ and proportional to $1/\sqrt{\eps_b}$ for $\eps_b < \eps_a$.

Lastly we turn to case (ii)  wherein
$\hat{\#c}_{\, a} = \hat{\#c}_{\, b} = \hat{\#z}$. The
  depolarization dyadics \r{D_dint} simplify to \c{WM_Needles_Pillboxes}
\begin{equation}
\=D_{\,\ell} = \frac{1}{ \eps_z }
  \hat{\#z} \, \hat{\#z}, \qquad \quad \le \ell = a, b \ri,
\end{equation}
and the corresponding solution to the dyadic Bruggeman equation
\r{Br}
\begin{equation} \l{eps_pa}
\left.
\begin{array}{l}
\eps_x = \eps_y = f_a \eps_a + f_b \eps_b
\vspace{8pt} \\
\eps_z =  \displaystyle{\frac{ \eps_a  \eps_b}{f_b \eps_a + f_a \eps_b }}
\end{array}
\right\}
\end{equation}
emerges.
We repeat the calculations of Fig.~\ref{fig6} but with $\hat{\#c}_{\, a} = \hat{\#c}_{\, b} = \hat{\#z}$. The corresponding plots of $\eps_y / \eps_z$ versus $\eps_b$ are presented in Fig.~\ref{fig7}. As in Fig.~\ref{fig6}, the magnitudes $\eps_y/\eps_z$ in Fig.~\ref{fig7}  deviate  from unity as $\eps_b$ deviates from 1.5 (the value of $\eps_a$), but the rates of growth of the $\eps_y/\eps_z$ curves in Fig.~\ref{fig7} are greater that those for the corresponding curves in Fig.~\ref{fig6}.
 For the range of $\eps_b$ and $f_a$ values considered in Fig.~\ref{fig7}, the magnitudes of  $\eps_y/ \eps_z$
 lie within the interval $\le 1, 37\ri$.

As previously, let us
  consider the  expressions for $\eps_y$ and $\eps_z$ given in Eqs.~\r{eps_pa} for the instance  $f_a = f_b = 0.5$. We find that
\begin{equation}
\frac{\eps_y}{\eps_z} = \frac{1}{4} \frac{\le \eps_a + \eps_b \ri^2}{\eps_a \eps_b} \to \infty \quad \mbox{as} \quad \eps_b \to \left\{ \begin{array}{l} 0 \\ \infty \end{array} \right. \quad \mbox{for fixed $\eps_a$.}
\end{equation}
That is, the degree of anisotropy, as gauged by  $\eps_y / \eps_z$, can increase without limit as $\eps_b$ increasingly deviates from $\eps_a$, and  the degree of anisotropy is proportional to $\eps_b$ for $\eps_b > \eps_a$ and proportional to $1/\eps_b$ for $\eps_b < \eps_a$.

\section{Closing remarks}

When a random mixture of two isotropic dielectric materials, one composed of oriented spheroidal particles of relative permittivity $\eps_a$ and the other composed of oriented spheroidal particles of relative permittivity $\eps_b$, is considered in the  long wavelength regime, the resulting HCM is  either an orthorhombic biaxial or a uniaxial dielectric material.
The degree of anisotropy exhibited by the HCM depends upon the eccentricity of the constituent spheroidal particles, and it is greatest when the alignments of the two populations of spheroids are the same. The greatest degrees of HCM anisotropy are achieved when the constituent particles are shaped as needles or discs. In these instances, explicit formulas for the HCM anisotropy may be derived from the dyadic Bruggeman equation \r{Br}. Using these formulas at fixed values of volume fraction and $\eps_a$, we find that the degrees of HCM anisotropy are proportional to $\sqrt{\eps_b}$ or $\eps_b$ for $\eps_b > \eps_a$,
and proportional to $1/\sqrt{\eps_b}$ or $1/\eps_b$ for $\eps_b < \eps_a$. Thus, in principle, there is no limit to degree of anisotropy that may be attained via homogenization. In practice, the degree of anisotropy would be limited by the available value of $\eps_b$ (and/or $\eps_a$). These findings may be helpful to those engaged in the development of anisotropic nanostructured composite materials for specific functions. For example, the described homogenization process may enable the very high degrees of anisotropy
which are required to create material analogues for various curved spacetime \c{Smolyaninov_NJP,Lu_JAP,ML_PRB,Spinning_string}
and quantum electrodynamical \c{QED}
scenarios to be attained.

Owing to the  electric--magnetic duality  intrinsic to the  Maxwell equations \c{EAB,PiO}, the findings presented herein apply equally well to magnetic properties. That is, by the homogenization of a random mixture of isotropic magnetic materials, distributed as oriented spheroidal particles, very high degrees of magnetic anisotropy may be attained.

%How eccentric does a spheroid need to be before it can be regarded as a needle/disc?

\section*{Appendix}

The  double integrals on the right side of Eqs.~\r{D_dint} yield the
 depolarization dyadics $\=D^{}_{\, \ell}$.
   Here we present  evaluations of these integrals.
   Let us begin with case (i), wherein the HCM is an orthorhombic biaxial dielectric material.
By symmetry considerations, the off-diagonal terms of $\=D^{}_{\, \ell}$ are null-valued; thus, we have   the diagonal form
\begin{equation} \l{D_diag}
\=D^{}_{\, \ell} = %\frac{1}{i \omega m_0}
\le \=U^{-1}_{\,\ell} \ri \.  \le \tilde{D}^{x}_{\ell} \, \hat{\#x} \, \hat{\#x}  + \tilde{D}^{y}_{\ell} \, \hat{\#y} \, \hat{\#y} +
\tilde{D}^{z}_{\ell} \, \hat{\#z} \, \hat{\#z}  \ri  \. \le \=U^{-1}_{\,\ell} \ri, \qquad \qquad \le \ell = a, b \ri.
\end{equation}

If we integrate the components $\tilde{D}^{x,y,z}_{\ell}$ first with respect to $\phi$ and then introduce the new variable $u = \cos \theta$, we find
\begin{eqnarray}
&&
\left. \l{D1}
\begin{array}{l}
\tilde{D}^{x}_\ell = \displaystyle{\frac{1}{\tilde{\eps}^{x}_\ell - \tilde{\eps}^{y}_\ell} \le 1 - \int^1_0 du \sqrt{ \frac{\tilde{\eps}^{y}_\ell + \le \tilde{\eps}^{z}_\ell - \tilde{\eps}^{y}_\ell \ri u^2}{\tilde{\eps}^{x}_\ell + \le \tilde{\eps}^{z}_\ell - \tilde{\eps}^{x}_\ell \ri u^2}} \ri} \,
\\[-2mm]
\\
\tilde{D}^{y}_\ell =  \displaystyle{\frac{1}{\tilde{\eps}^{y}_\ell - \tilde{\eps}^{x}_\ell}
\le 1 - \int^1_0 du \sqrt{ \frac{\tilde{\eps}^{x}_\ell + \le \tilde{\eps}^{z}_\ell - \tilde{\eps}^{x}_\ell \ri u^2}{\tilde{\eps}^{y}_\ell
+ \le \tilde{\eps}^{z}_\ell - \tilde{\eps}^{y}_\ell \ri u^2}} \ri} \,
\\[-2mm]
\\
\tilde{D}^{z}_\ell = \displaystyle{ \int^1_0 du
 \frac{u^2}{\sqrt{ \les \tilde{\eps}^{x}_\ell + \le \tilde{\eps}^{z}_\ell - \tilde{\eps}^{x}_\ell \ri u^2\ris
 \les \tilde{\eps}^{y}_\ell + \le \tilde{\eps}^{z}_\ell - \tilde{\eps}^{y}_\ell \ri u^2\ris}}}  \,
\end{array}
\right\},
 \quad \le \ell = a, b \ri,
% \nonumber \\ &&  \hspace{75mm} ( \tilde{m} = \eps, \mu ).
\end{eqnarray}
wherein
\begin{equation}
\tilde{\eps}^{n}_\ell = \frac{\eps^{}_n}{\le U^n_\ell \ri^2}, \qquad \qquad  \le \ell = a, b;  n = x, y, z  \ri ,
\end{equation}
with $U^n_\ell$ being the diagonal components of the shape dyadics $\=U_{a,b}$, i.e.,
\begin{equation}
\=U_\ell \equiv U^{x}_{\ell} \, \hat{\#x} \, \hat{\#x} + U^{y}_{\ell} \, \hat{\#y} \, \hat{\#y} + U^{z}_{\ell} \, \hat{\#z} \, \hat{\#z},  \qquad \le \ell = a, b \ri.
\end{equation}

The terms on the right sides in Eqs.~\r{D1} are expressible in terms of incomplete elliptic integrals; the form of these elliptic integral expressions depends upon the relative sizes of $\tilde{\eps}^{x}_\ell$, $\tilde{\eps}^{y}_\ell$ and $\tilde{\eps}^{z}_\ell$. For $\tilde{\eps}^{x,y,z}_\ell > 0 $ we find
\begin{equation} \l{Dx}
\tilde{D}^{x}_\ell =
\left\{
\begin{array}{l}
 \displaystyle{\frac{1+ i \sqrt{\frac{\tilde{\eps}^{y}_\ell}{\tilde{\eps}^{z}_\ell - \tilde{\eps}^{x}_\ell}} E \le i \sinh^{-1} \sqrt{\frac{\tilde{\eps}^{z}_\ell}{\tilde{\eps}^{x}_\ell}-1}, \sqrt{\frac{\tilde{\eps}^{x}_\ell \le \tilde{\eps}^{y}_\ell-\tilde{\eps}^{z}_\ell \ri}{\tilde{\eps}^{y}_\ell \le \tilde{\eps}^{x}_\ell - \tilde{\eps}^{z}_\ell \ri}} \ri}{\tilde{\eps}^{x}_\ell - \tilde{\eps}^{y}_\ell}}
, \qquad
\begin{array}{l}
 \tilde{\eps}^{z}_\ell > \tilde{\eps}^{y}_\ell > \tilde{\eps}^{x}_\ell, \\ \tilde{\eps}^{y}_\ell > \tilde{\eps}^{z}_\ell > \tilde{\eps}^{x}_\ell, \\ \tilde{\eps}^{z}_\ell > \tilde{\eps}^{x}_\ell > \tilde{\eps}^{y}_\ell;
 \end{array} \vspace{7mm}
\\
\displaystyle{\frac{1 - \sqrt{\frac{\tilde{\eps}^{y}_\ell}{\tilde{\eps}^{x}_\ell - \tilde{\eps}^{z}_\ell}} E \le  \sec^{-1} \sqrt{\frac{\tilde{\eps}^{x}_\ell}{\tilde{\eps}^{z}_\ell}}, \sqrt{\frac{\tilde{\eps}^{x}_\ell \le \tilde{\eps}^{y}_\ell-\tilde{\eps}^{z}_\ell \ri}{\tilde{\eps}^{y}_\ell \le \tilde{\eps}^{x}_\ell - \tilde{\eps}^{z}_\ell \ri}} \ri}{\tilde{\eps}^{x}_\ell - \tilde{\eps}^{y}_\ell}}
, \qquad
\begin{array}{l}
 \tilde{\eps}^{y}_\ell > \tilde{\eps}^{x}_\ell > \tilde{\eps}^{z}_\ell, \\ \tilde{\eps}^{x}_\ell > \tilde{\eps}^{z}_\ell > \tilde{\eps}^{y}_\ell, \\ \tilde{\eps}^{x}_\ell > \tilde{\eps}^{y}_\ell > \tilde{\eps}^{z}_\ell;
 \end{array}
 \end{array}
\right.
\end{equation}
\begin{equation}
\tilde{D}^{y}_\ell =
\left\{
\begin{array}{l}
 \displaystyle{\frac{-1- i \sqrt{\frac{\tilde{\eps}^{x}_\ell}{\tilde{\eps}^{z}_\ell - \tilde{\eps}^{y}_\ell}} E \le i \sinh^{-1} \sqrt{\frac{\tilde{\eps}^{z}_\ell}{\tilde{\eps}^{y}_\ell}-1}, \sqrt{\frac{\tilde{\eps}^{y}_\ell \le \tilde{\eps}^{x}_\ell-\tilde{\eps}^{z}_\ell \ri}{\tilde{\eps}^{x}_\ell \le \tilde{\eps}^{y}_\ell - \tilde{\eps}^{z}_\ell \ri}} \ri}{\tilde{\eps}^{x}_\ell - \tilde{\eps}^{y}_\ell}}
, \qquad
\begin{array}{l}
 \tilde{\eps}^{z}_\ell > \tilde{\eps}^{y}_\ell > \tilde{\eps}^{x}_\ell, \\ \tilde{\eps}^{x}_\ell > \tilde{\eps}^{z}_\ell > \tilde{\eps}^{y}_\ell, \\ \tilde{\eps}^{z}_\ell > \tilde{\eps}^{x}_\ell > \tilde{\eps}^{y}_\ell;
 \end{array} \vspace{7mm}
\\
\displaystyle{\frac{- 1 + \sqrt{\frac{\tilde{\eps}^{x}_\ell}{\tilde{\eps}^{y}_\ell - \tilde{\eps}^{z}_\ell}} E \le  \sec^{-1} \sqrt{\frac{\tilde{\eps}^{y}_\ell}{\tilde{\eps}^{z}_\ell}}, \sqrt{\frac{\tilde{\eps}^{y}_\ell \le \tilde{\eps}^{x}_\ell-\tilde{\eps}^{z}_\ell \ri}{\tilde{\eps}^{x}_\ell \le \tilde{\eps}^{y}_\ell - \tilde{\eps}^{z}_\ell \ri}} \ri}{\tilde{\eps}^{x}_\ell - \tilde{\eps}^{y}_\ell}}
, \qquad
\begin{array}{l}
 \tilde{\eps}^{y}_\ell > \tilde{\eps}^{x}_\ell > \tilde{\eps}^{z}_\ell, \\ \tilde{\eps}^{y}_\ell > \tilde{\eps}^{z}_\ell > \tilde{\eps}^{x}_\ell, \\ \tilde{\eps}^{x}_\ell > \tilde{\eps}^{y}_\ell > \tilde{\eps}^{z}_\ell;
 \end{array}
 \end{array}
\right.
\end{equation}
and
\begin{equation} \l{Dz}
\tilde{D}^{z}_\ell =
\left\{
\begin{array}{l}
 \displaystyle{\frac{ F \le \sec^{-1} \sqrt{\frac{\tilde{\eps}^{x}_\ell}{\tilde{\eps}^{z}_\ell}}, \sqrt{\frac{\tilde{\eps}^{x}_\ell \le \tilde{\eps}^{y}_\ell - \tilde{\eps}^{z}_\ell \ri}{\tilde{\eps}^{y}_\ell \le \tilde{\eps}^{x}_\ell - \tilde{\eps}^{z}_\ell \ri}} \ri  - \, E \le \sec^{-1} \sqrt{\frac{\tilde{\eps}^{x}_\ell}{\tilde{\eps}^{z}_\ell}}, \sqrt{\frac{\tilde{\eps}^{x}_\ell \le \tilde{\eps}^{y}_\ell - \tilde{\eps}^{z}_\ell \ri}{\tilde{\eps}^{y}_\ell \le \tilde{\eps}^{x}_\ell - \tilde{\eps}^{z}_\ell \ri}} \ri }{\le \tilde{\eps}^{y}_\ell - \tilde{\eps}^{z}_\ell \ri \frac{\sqrt{ \tilde{\eps}^{x}_\ell - \tilde{\eps}^{z}_\ell }}{\sqrt{\tilde{\eps}^{y}_\ell} }      }}
, \qquad
\begin{array}{l}
 \tilde{\eps}^{y}_\ell > \tilde{\eps}^{x}_\ell > \tilde{\eps}^{z}_\ell, \\ \tilde{\eps}^{x}_\ell > \tilde{\eps}^{z}_\ell > \tilde{\eps}^{y}_\ell, \\ \tilde{\eps}^{x}_\ell > \tilde{\eps}^{y}_\ell > \tilde{\eps}^{z}_\ell;
 \end{array} \vspace{7mm}
\\
 \displaystyle{\frac{  F \le i \sinh^{-1} \sqrt{\frac{\tilde{\eps}^{z}_\ell - \tilde{\eps}^{y}_\ell}{\tilde{\eps}^{y}_\ell}}, \sqrt{\frac{\tilde{\eps}^{y}_\ell \le \tilde{\eps}^{z}_\ell - \tilde{\eps}^{x}_\ell \ri}{\tilde{\eps}^{x}_\ell \le \tilde{\eps}^{z}_\ell - \tilde{\eps}^{y}_\ell \ri}} \ri  - \, E   \le  i \sinh^{-1} \sqrt{\frac{\tilde{\eps}^{z}_\ell - \tilde{\eps}^{y}_\ell}{\tilde{\eps}^{y}_\ell}}, \sqrt{\frac{\tilde{\eps}^{y}_\ell \le \tilde{\eps}^{z}_\ell - \tilde{\eps}^{x}_\ell \ri}{\tilde{\eps}^{x}_\ell \le \tilde{\eps}^{z}_\ell - \tilde{\eps}^{y}_\ell \ri}} \ri}{\le \tilde{\eps}^{z}_\ell - \tilde{\eps}^{x}_\ell \ri \frac{\sqrt{ \tilde{\eps}^{z}_\ell - \tilde{\eps}^{y}_\ell}}{ i \sqrt{\tilde{\eps}^{x}_\ell}}         }}
, \qquad
\begin{array}{l}
 \tilde{\eps}^{z}_\ell > \tilde{\eps}^{y}_\ell > \tilde{\eps}^{x}_\ell, \\ \tilde{\eps}^{z}_\ell > \tilde{\eps}^{x}_\ell > \tilde{\eps}^{y}_\ell;
 \end{array}
 \vspace{7mm}
\\
 \displaystyle{\frac{  E \le \sin^{-1} \sqrt{\frac{\tilde{\eps}^{z}_\ell - \tilde{\eps}^{x}_\ell}{\tilde{\eps}^{x}_\ell}}, \sqrt{\frac{\tilde{\eps}^{x}_\ell \le \tilde{\eps}^{y}_\ell - \tilde{\eps}^{z}_\ell \ri}{\tilde{\eps}^{y}_\ell \le \tilde{\eps}^{z}_\ell - \tilde{\eps}^{x}_\ell \ri}} \ri  - \, F   \le \sin^{-1} \sqrt{\frac{\tilde{\eps}^{z}_\ell - \tilde{\eps}^{x}_\ell}{\tilde{\eps}^{x}_\ell}}, \sqrt{\frac{\tilde{\eps}^{x}_\ell \le \tilde{\eps}^{y}_\ell - \tilde{\eps}^{z}_\ell \ri}{\tilde{\eps}^{y}_\ell \le \tilde{\eps}^{z}_\ell - \tilde{\eps}^{x}_\ell \ri}} \ri  }{
 \le \tilde{\eps}^{z}_\ell - \tilde{\eps}^{y}_\ell \ri
\frac{ \sqrt{ \tilde{\eps}^{z}_\ell - \tilde{\eps}^{x}_\ell }}{ \sqrt{\tilde{\eps}^{y}_\ell} }   }}
, \qquad
\begin{array}{l}
 \tilde{\eps}^{y}_\ell > \tilde{\eps}^{z}_\ell > \tilde{\eps}^{x}_\ell.
 \end{array}
 \end{array}
\right.
\end{equation}
The quantities $F(\varphi, k)$ and $E(\varphi, k)$ herein are the incomplete elliptic integrals of the first and second kind, respectively, as defined by \c{Grad}
\begin{equation}
\left.
\begin{array}{l}
F(\varphi, k) = \displaystyle{ \int^\varphi_0 \frac{d t}{\sqrt{1- k^2 \sin^2 t}}} \vspace{6pt} \\
E(\varphi, k) = \displaystyle{ \int^\varphi_0 \sqrt{1- k^2 \sin^2 t} \; d t }
\end{array}
\right\},
\end{equation}
with $\varphi$ being the \emph{amplitude}.
For compact representation, imaginary-valued amplitudes are used in Eqs.~\r{Dx}--\r{Dz}, but all depolarization dyadic components
 herein are real-valued. Standard elliptic integral identities \c{AS} may be used to re-express Eqs.~\r{Dx}--\r{Dz} in terms of real-valued amplitudes. The expressions~\r{Dx}--\r{Dz} represent a generalization of the corresponding results derived by Weiglhofer for spherical particles embedded in a biaxial dielectric material \c{W98}.

Now, we turn to case (ii)  wherein the HCM is a uniaxial dielectric material. The depolarization dyadic retains the diagonal form \r{D_diag} but here $\tilde{D}^{x}_{\ell} = \tilde{D}^{y}_{\ell} $. The integrals on the right sides of Eqs.~\r{D1} may be evaluated as
\begin{eqnarray}
\tilde{D}^{x}_{\ell} &=& \frac{U^2_x}{ \eps^{}_x} \, \Gamma_x (\nu),  \l{Dxi}\\
\tilde{D}^{z}_{\ell} &=& \frac{U^2_z \nu}{ \eps^{}_z } \, \Gamma_z ( \nu ),
\l{Dzi}
\end{eqnarray}
wherein the terms
\begin{eqnarray}
\Gamma_x( \nu )&=& \left\{
\begin{array}{lcr}
\displaystyle{ \frac{1}{2} \le  \frac{1}{1-\nu }-
  \frac{ \nu \sinh^{-1} \sqrt{\frac{1
-\nu}{\nu} }}{\le 1 - \nu  \ri^{\frac{3}{2}}} \ri } &&
\mbox{for} \;\; 0 < \nu < 1
\\ & & \\
\displaystyle{\frac{1}{2} \le \frac{\nu \sec^{-1} \sqrt{\nu} }
{\le \nu - 1 \ri^{\frac{3}{2}}} - \frac{1}{\nu - 1}  \ri }& &
\mbox{for} \;\; \nu > 1
\end{array}
\right.,
\\
\Gamma_z (\nu )&=& \left\{
\begin{array}{lcr}
\displaystyle{
  \frac{\sinh^{-1} \sqrt{\frac{1
-\nu}{\nu} }}{\le 1 - \nu  \ri^{\frac{3}{2}}} -
\frac{1}{1-\nu }} && \hspace{14mm} \mbox{for} \;\; 0 < \nu < 1
\\ & & \\
\displaystyle{ \frac{1}{\nu - 1} - \frac{\sec^{-1} \sqrt{\nu}
} {\le \nu - 1 \ri^{\frac{3}{2}}}}& & \mbox{for} \;\; \nu > 1
\end{array}
\right., \\
\end{eqnarray}
with  the scalar parameter
\begin{equation}
\nu = \frac{U^2_x \eps^{}_z}{U^2_z
\eps^{}_x}.
\end{equation}
 The anomalous case $\nu < 0$, which corresponds to a
hyperbolic HCM \c{MLD2005}, is excluded from our consideration here.

\newpage

\begin{figure}[!ht]
\centering
\includegraphics[width=2.5in]{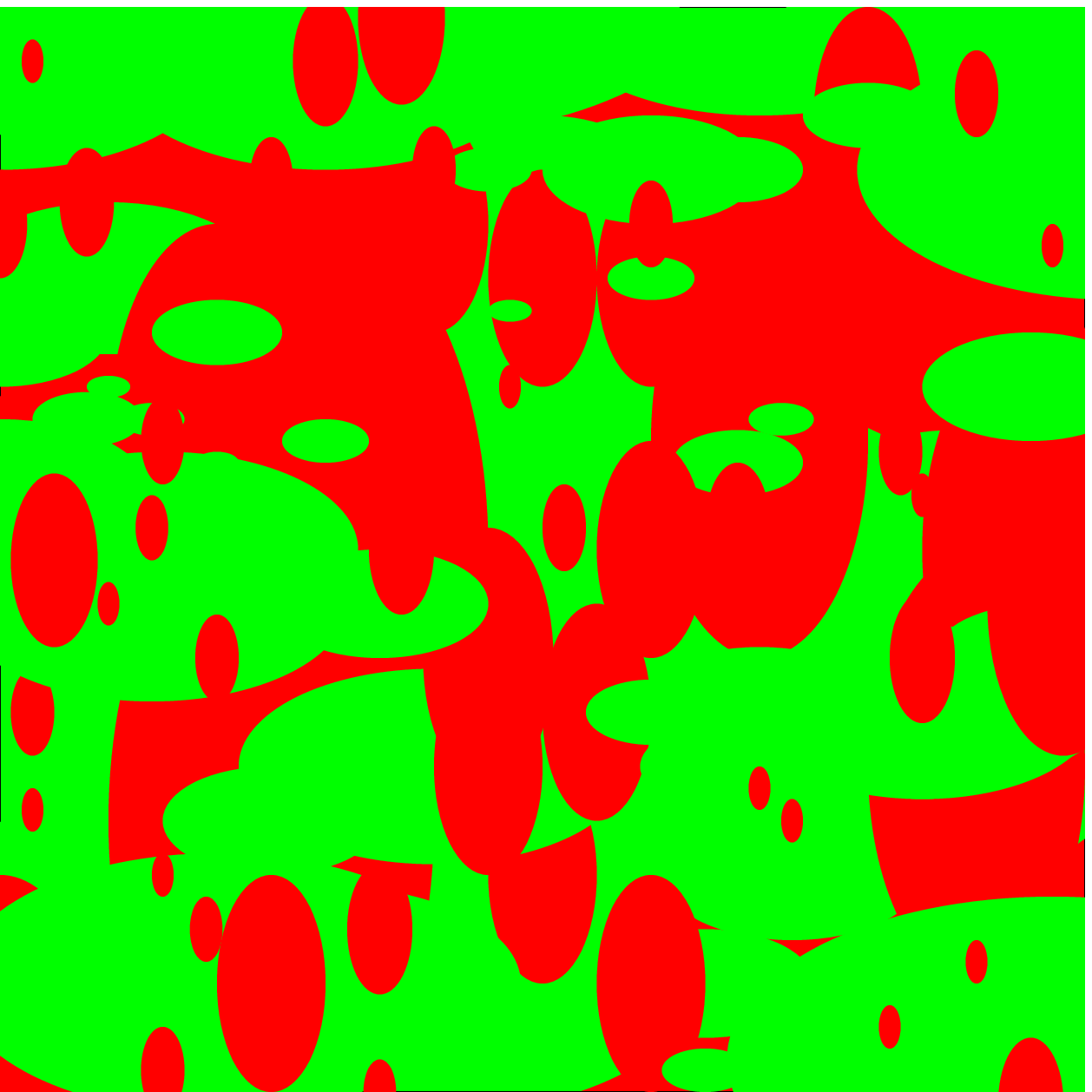}\hspace{10mm}
\includegraphics[width=2.5in]{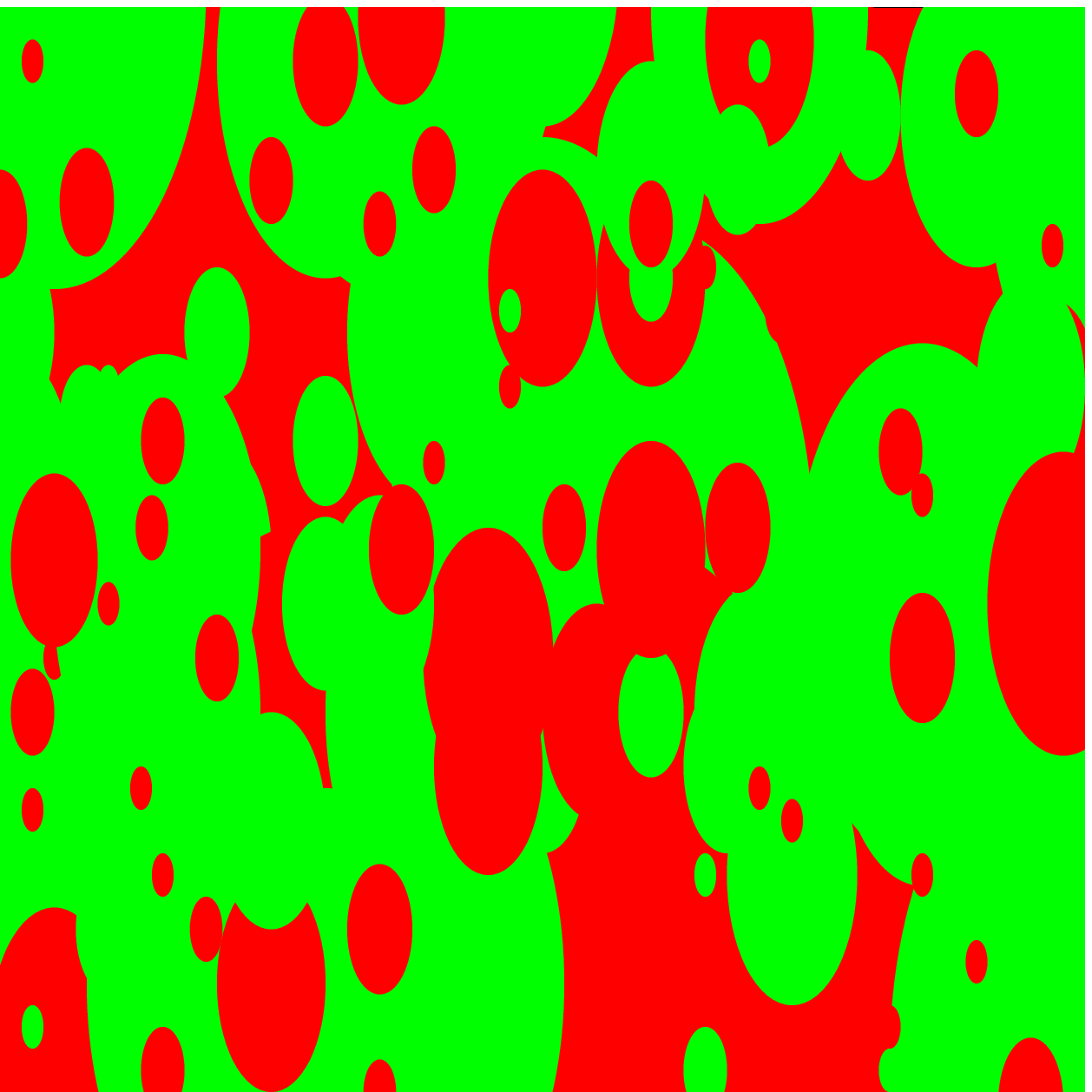}
 \caption{Two schematic representations of randomly-mixed component material $a$ and $b$ spheroids. The component material $a$ spheroids all have the same orientation and the  component material $b$ spheroids all have the same orientation; we consider cases wherein these two orientations are mutually perpendicular  (left) and are the same (right).  \l{fig1}}
\end{figure}

\newpage

\begin{figure}[!ht]
\centering
\includegraphics[width=2.5in]{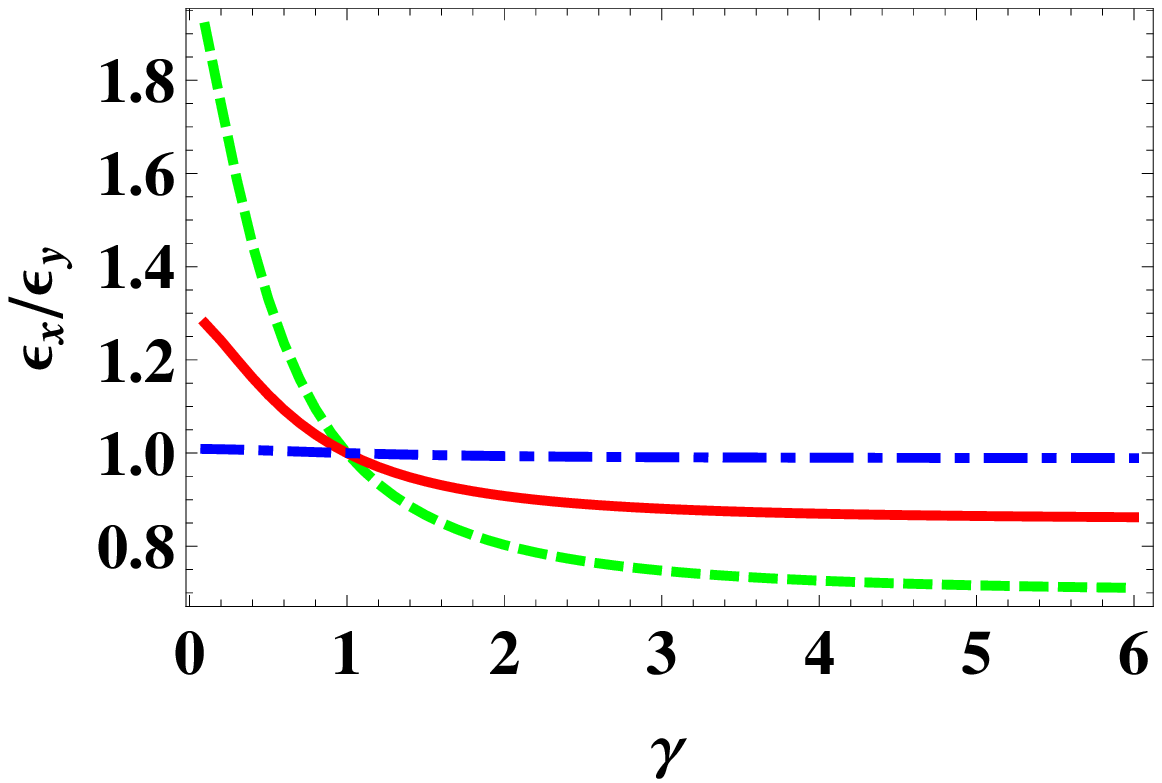}\hspace{10mm}
\includegraphics[width=2.5in]{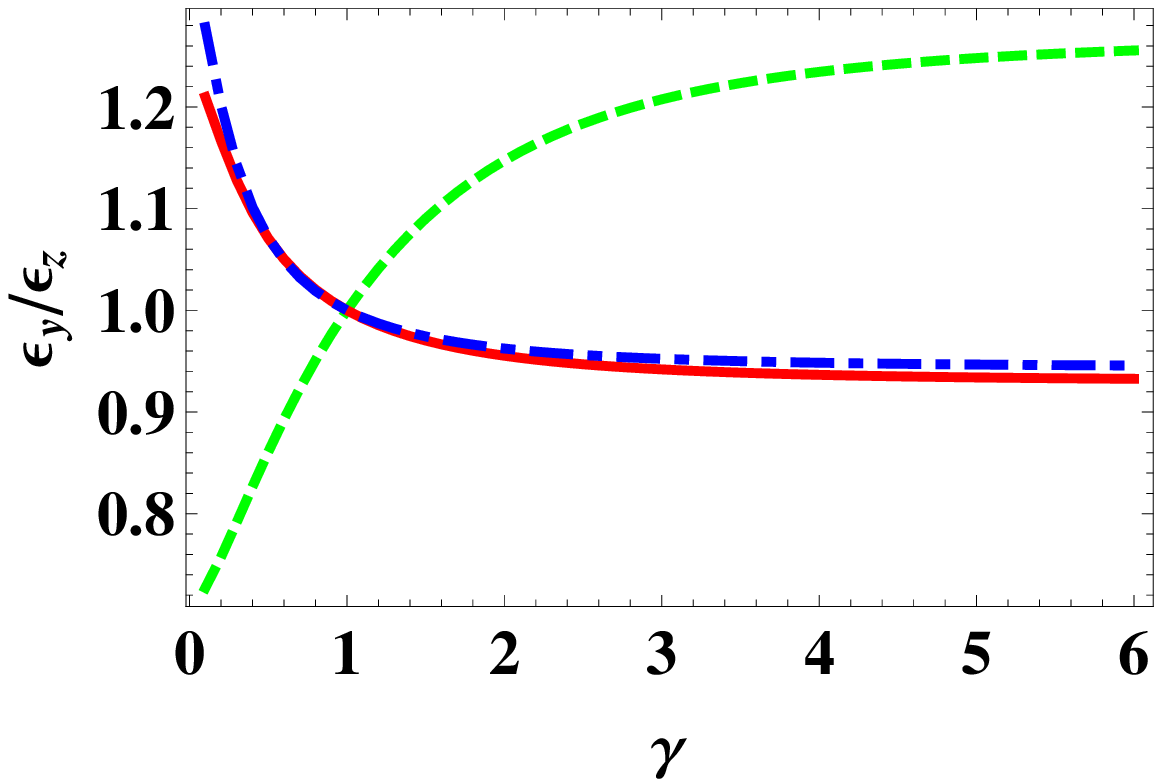}
\hspace{10mm}
\includegraphics[width=2.5in]{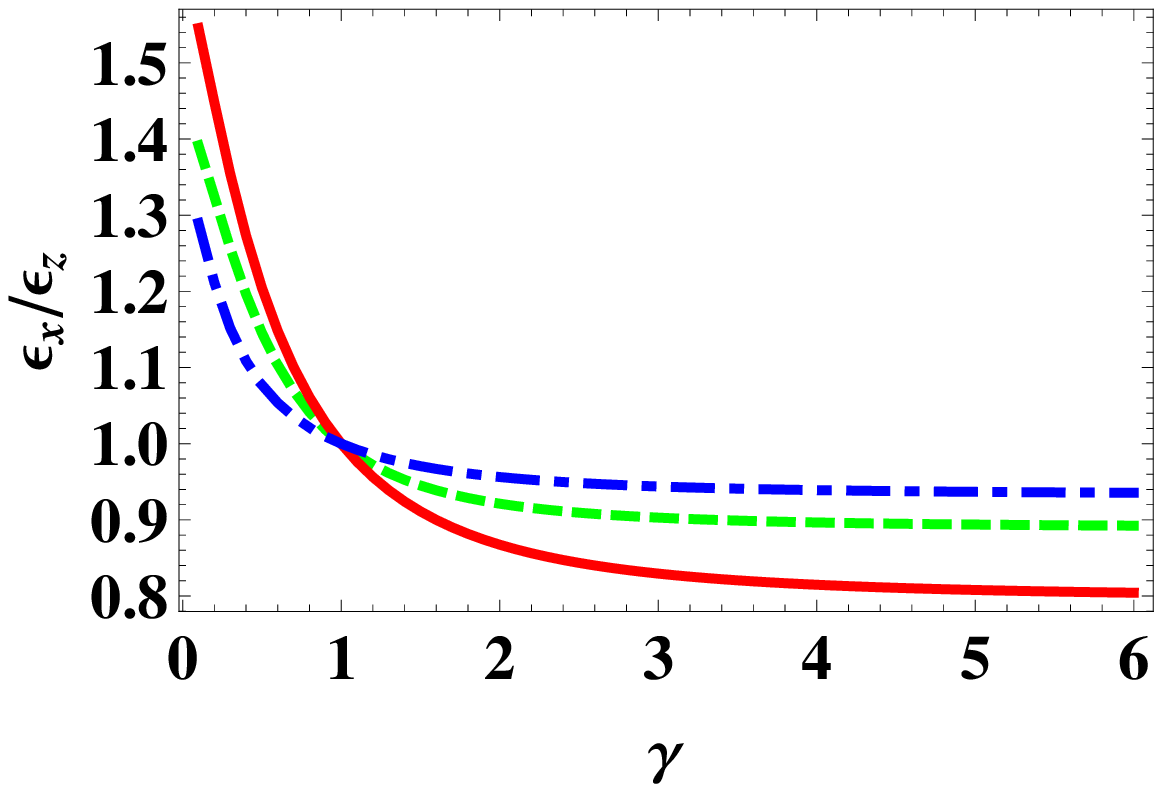}
 \caption{$\eps_x / \eps_y$, $\eps_y / \eps_z$ and $\eps_x / \eps_z$ plotted versus the eccentricity parameter $\gamma \in \le 0.1, 6 \ri$ for volume fractions $f_a = 0.7$ (green, dashed curves), 0.4 (red, solid curves) and 0.1 (blue, broken dashed curves). The symmetry axis  of the component $a$ spheroids is parallel to the $z$ coordinate axis whereas the symmetry axis  of the component $b$ spheroids is  parallel to the $y$ coordinate axis.
  \l{fig2}}
\end{figure}

\begin{figure}[!ht]
\centering
\includegraphics[width=2.5in]{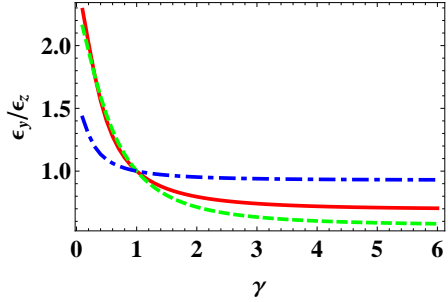}
 \caption{$\eps_y / \eps_z$ plotted versus the eccentricity parameter $\gamma \in \le 0.1, 6 \ri$ for volume fractions $f_a = 0.7$ (green, dashed curve), 0.4 (red, solid curve) and 0.1 (blue, broken dashed curve). The symmetry axes of the component  $a$ and $b$ spheroids are parallel to the $z$ coordinate axis.
  \l{fig3}}
\end{figure}

\newpage

\begin{figure}[!ht]
\centering
\includegraphics[width=2.5in]{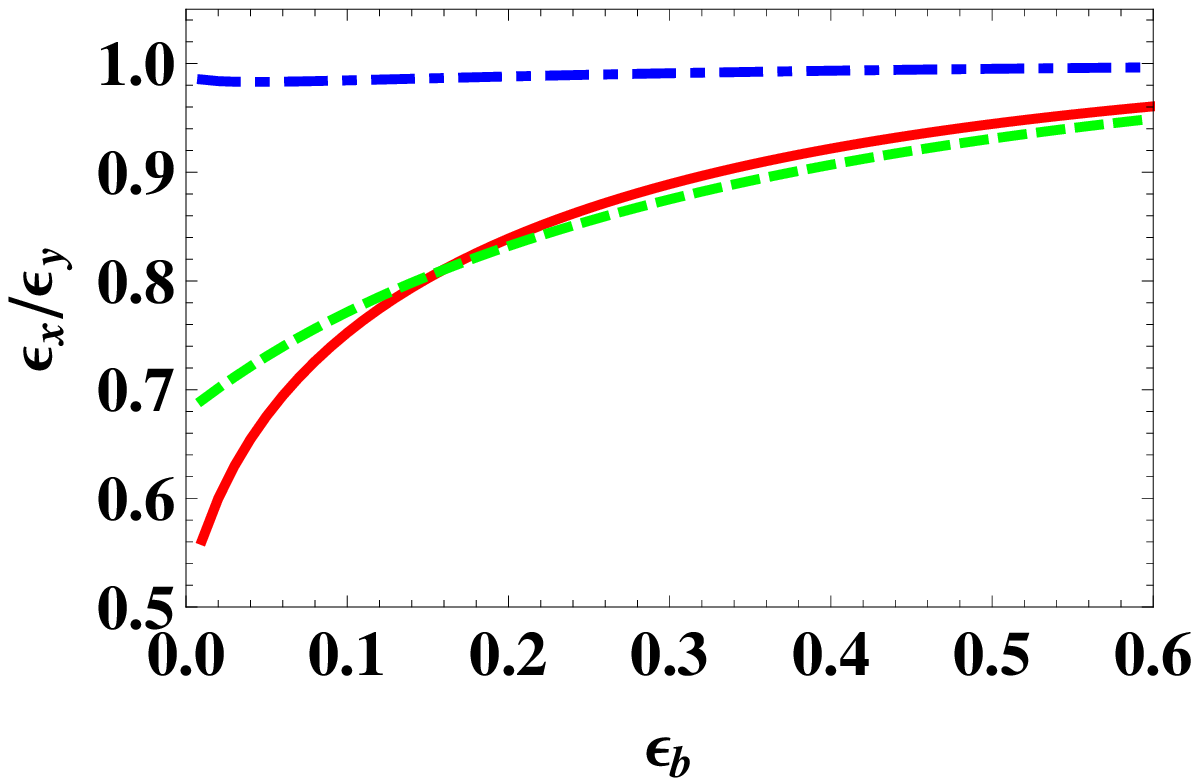}\hspace{10mm}
\includegraphics[width=2.5in]{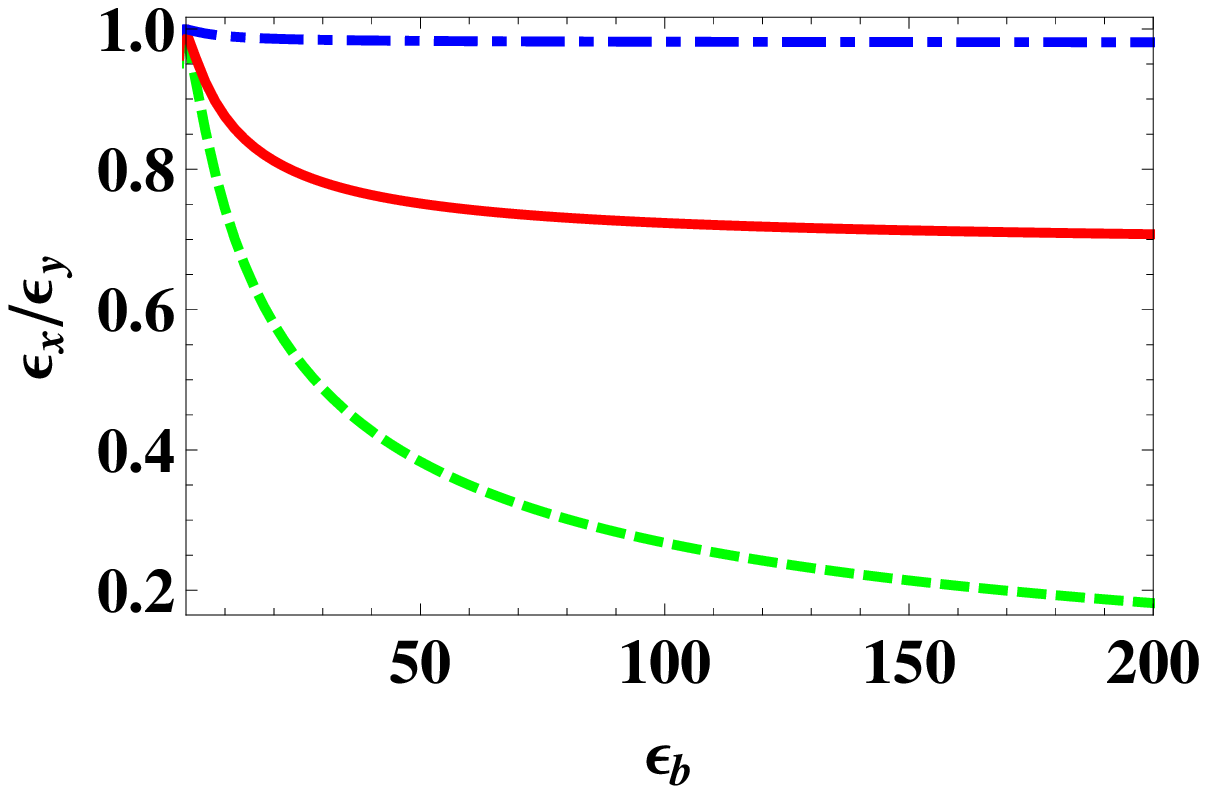}\\
\includegraphics[width=2.5in]{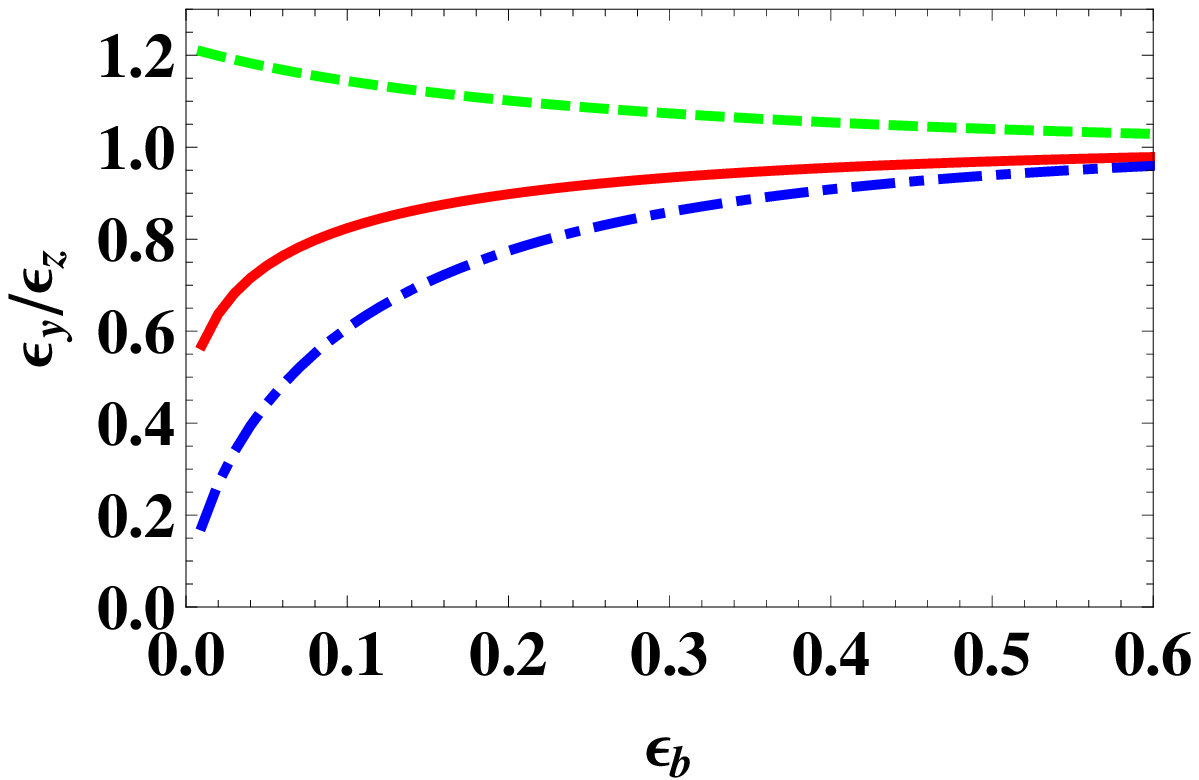}\hspace{10mm}
\includegraphics[width=2.5in]{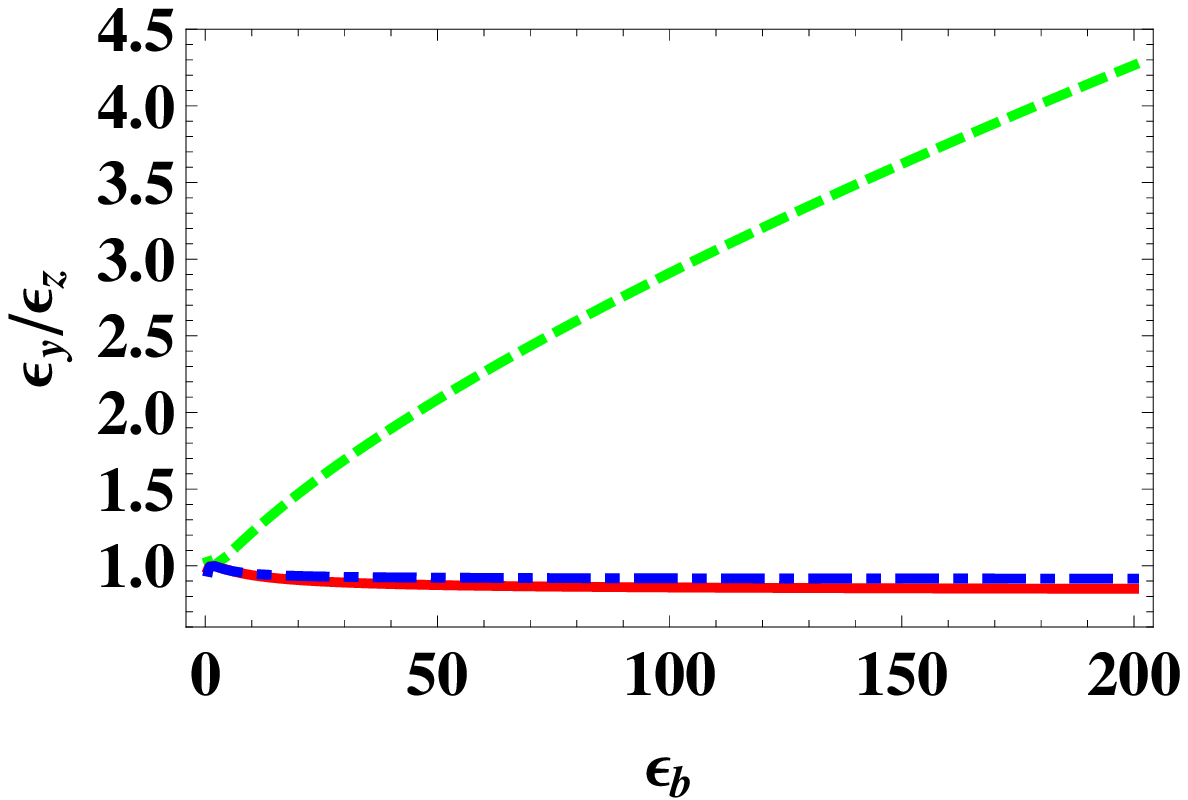}\\
\includegraphics[width=2.5in]{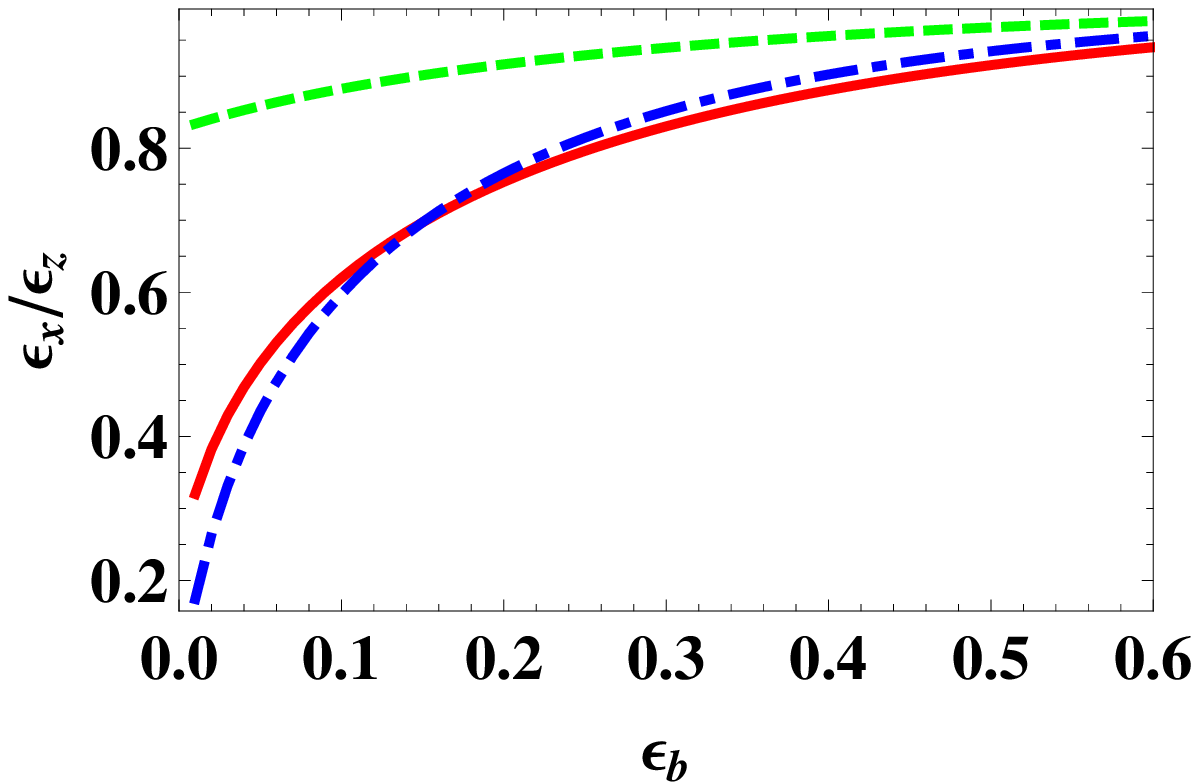}\hspace{10mm}
\includegraphics[width=2.5in]{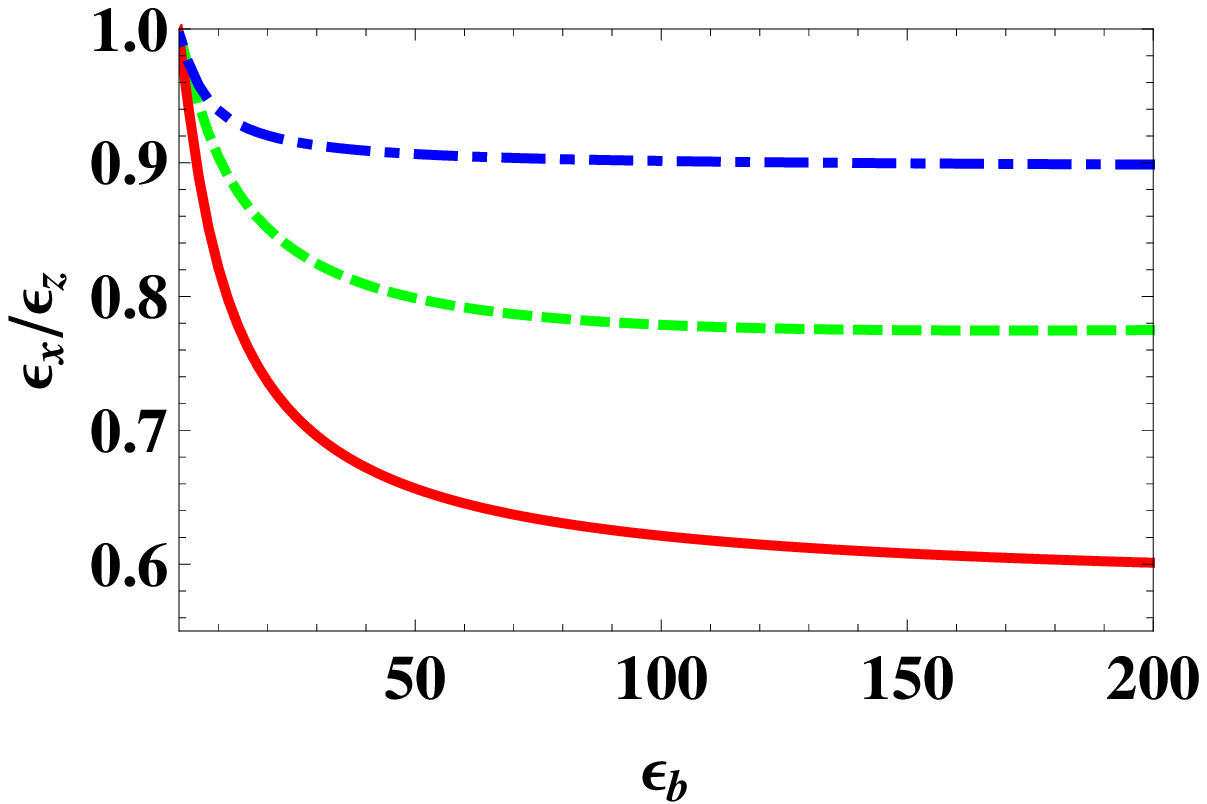}\\
 \caption{$\eps_x / \eps_y$, $\eps_y / \eps_z$ and $\eps_x / \eps_z$  plotted versus  $\eps_b \in \le 0.01, 0.6 \ri$ (left) and $\le  0.6, 200 \ri$ (right) for volume fractions $f_a = 0.7$ (green, dashed curves), 0.4 (red, solid curves) and 0.1 (blue, broken dashed curves). The component $a$ needles are  parallel to the $z$ coordinate axis whereas the component $b$ needles are  parallel to the $y$ coordinate axis.
  \l{fig4}}
\end{figure}

\newpage

\begin{figure}[!ht]
\centering
\includegraphics[width=2.5in]{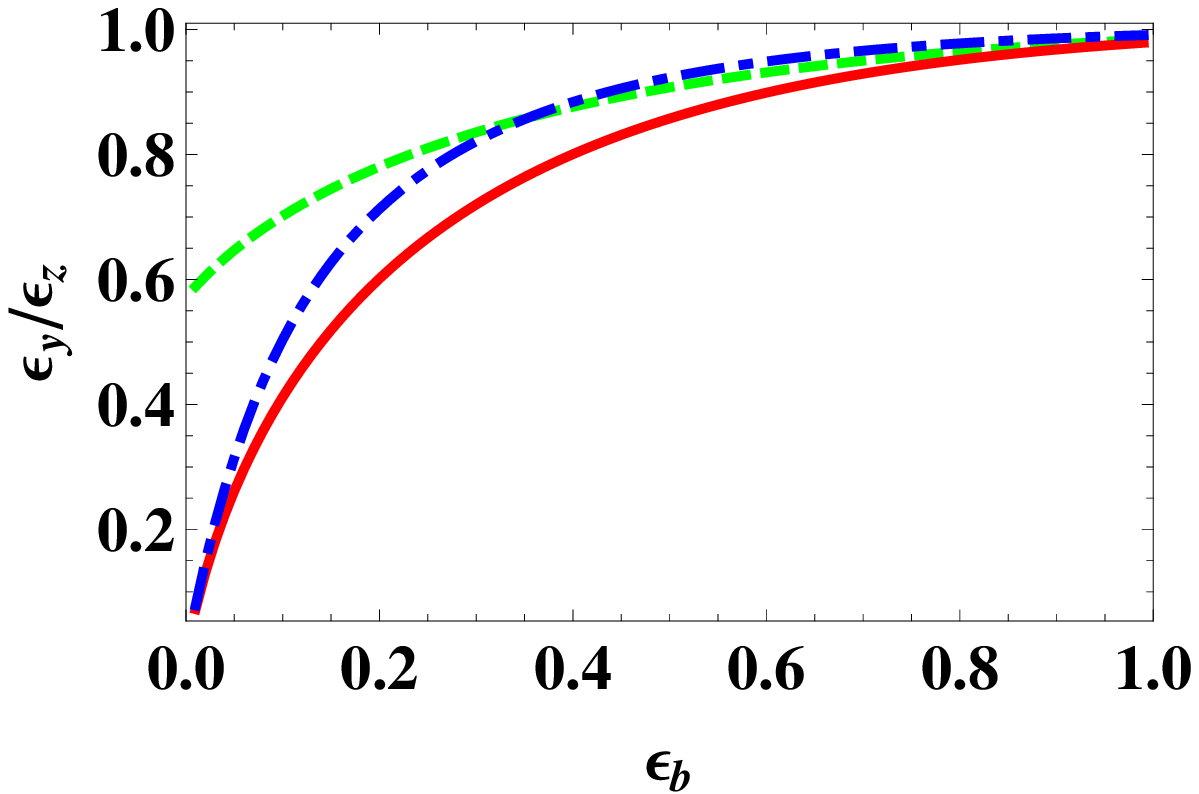}\hspace{10mm}
\includegraphics[width=2.5in]{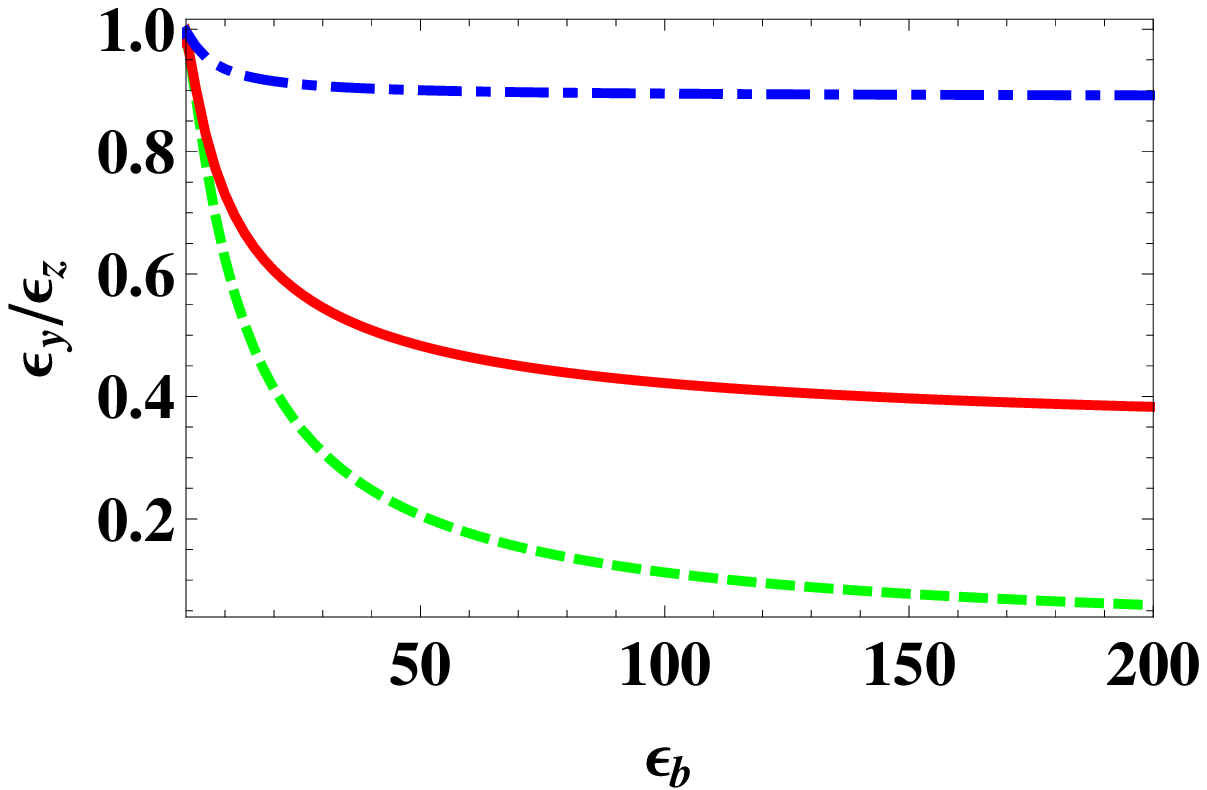}
 \caption{$\eps_y / \eps_z$ plotted versus  $\eps_b \in \le 0.01, 1 \ri$ (left) and $\le  1, 200 \ri$ (right) for volume fractions $f_a = 0.7$ (green, dashed curves), 0.4 (red, solid curves) and 0.1 (blue, broken dashed curves). The component  $a$ and $b$ needles are both  parallel to the $z$ coordinate axis.
  \l{fig5}}
\end{figure}

\newpage

\begin{figure}[!ht]
\centering
\includegraphics[width=2.5in]{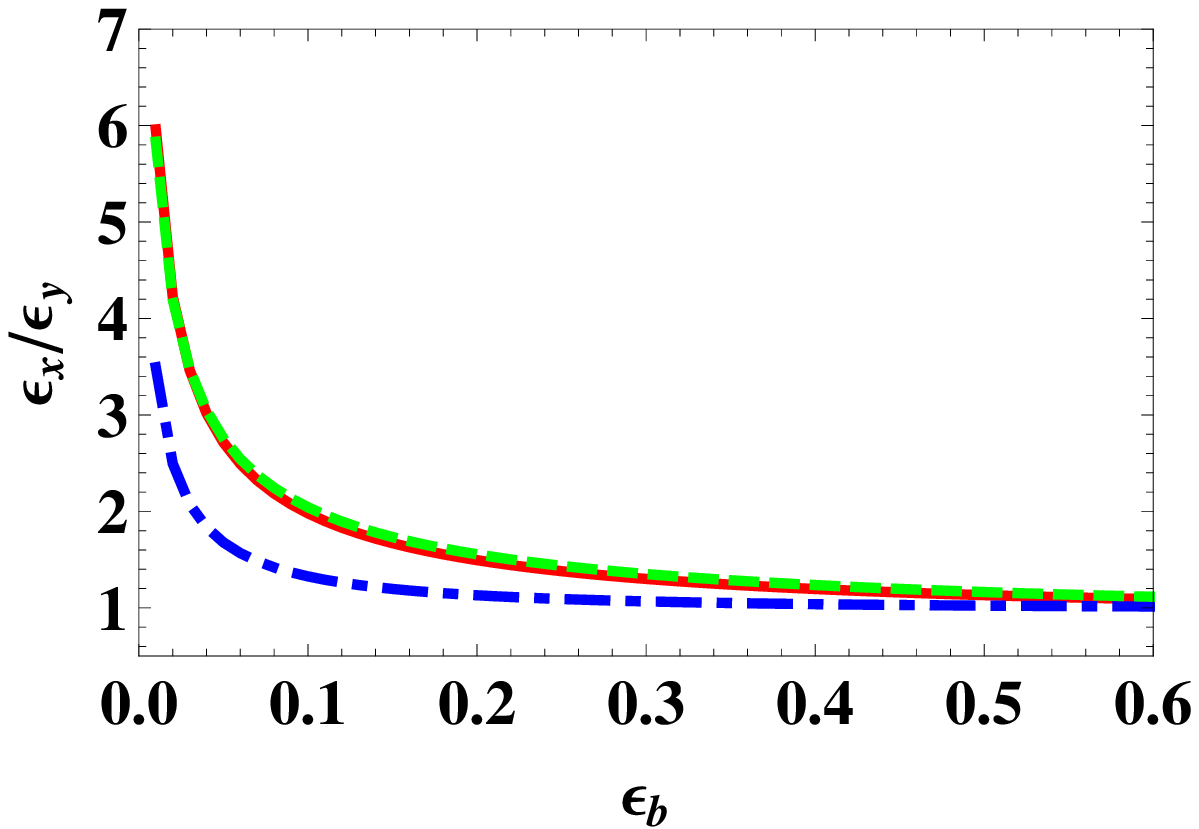}\hspace{10mm}
\includegraphics[width=2.5in]{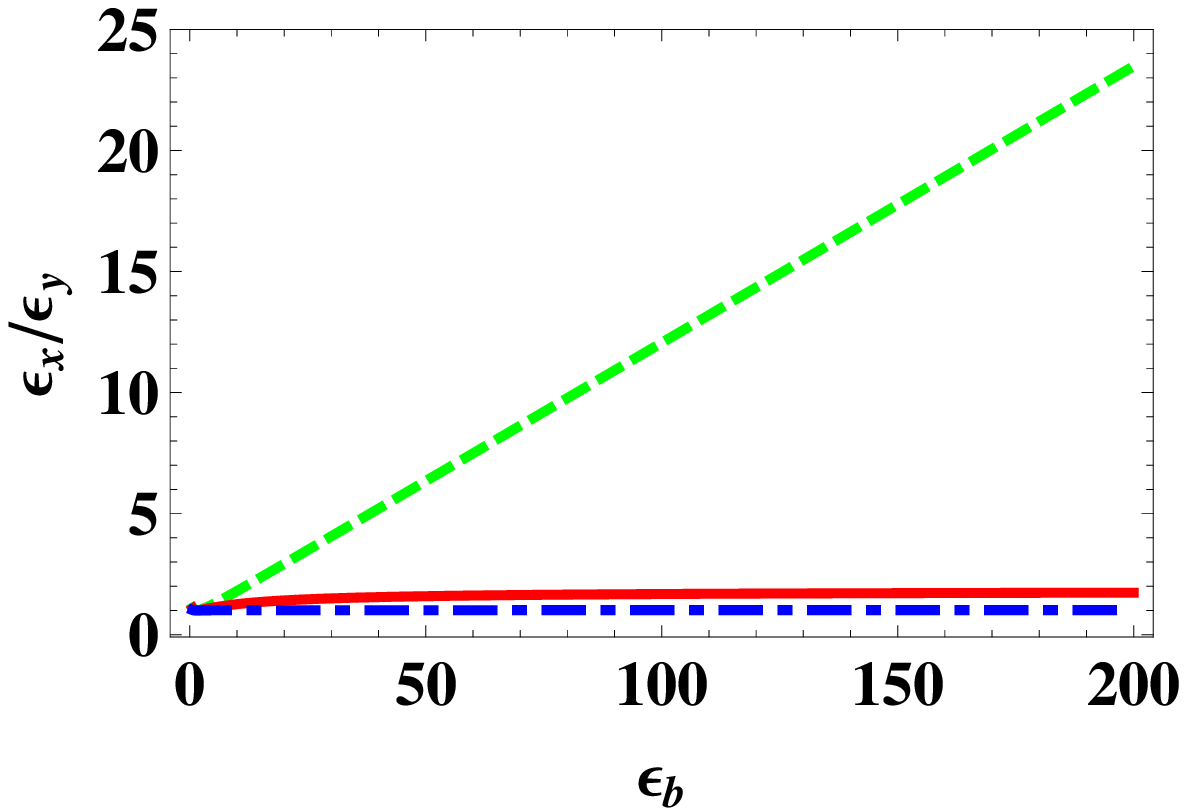}\\
\includegraphics[width=2.5in]{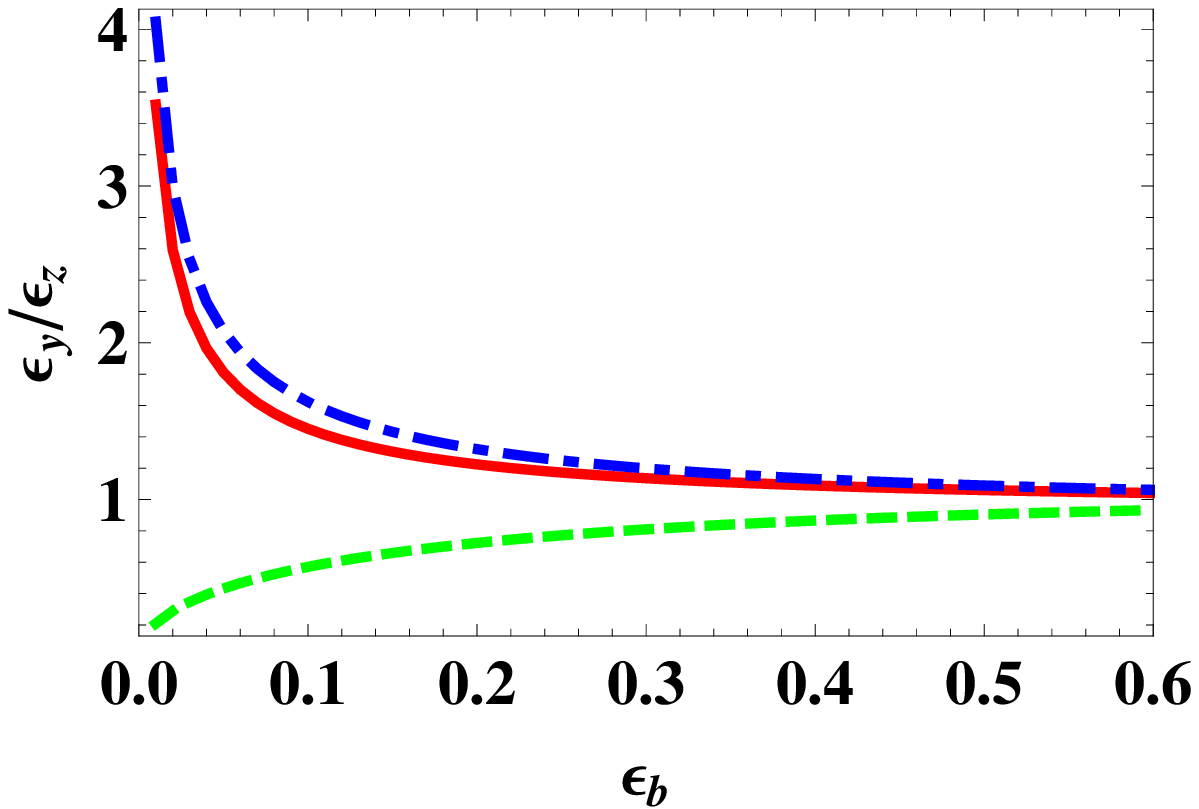}\hspace{10mm}
\includegraphics[width=2.5in]{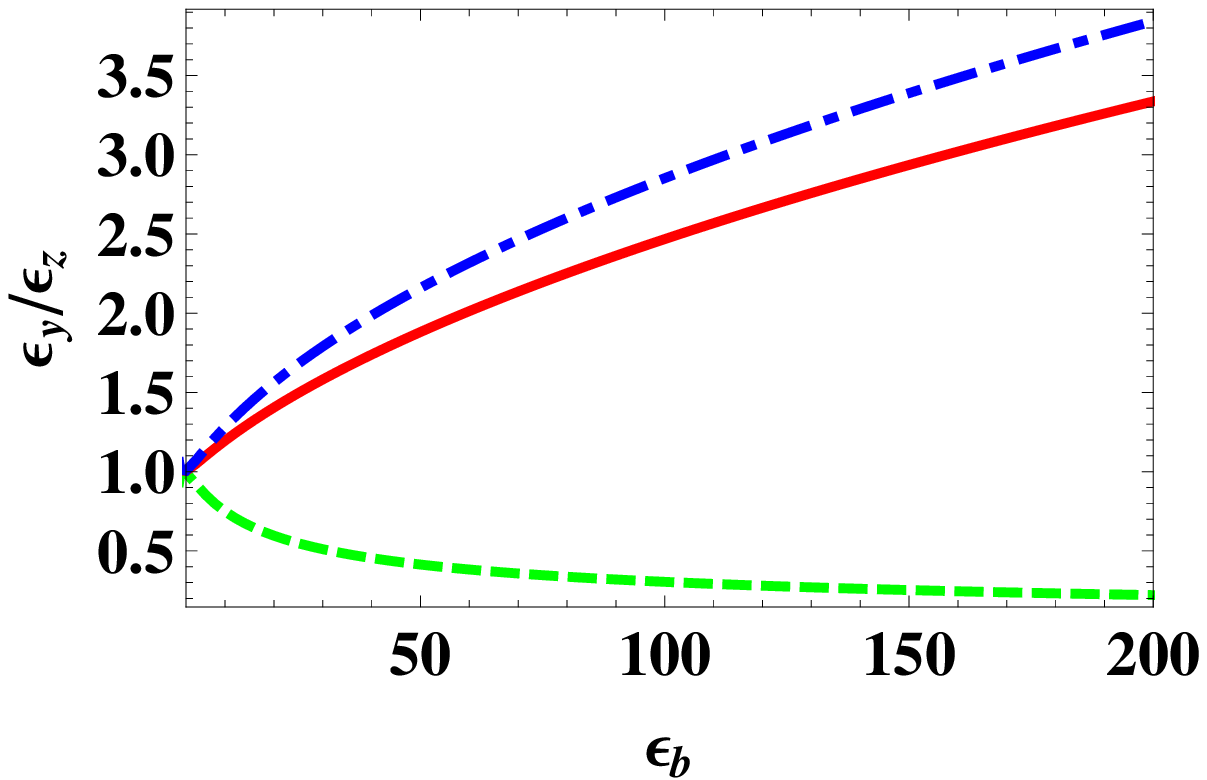}\\
\includegraphics[width=2.5in]{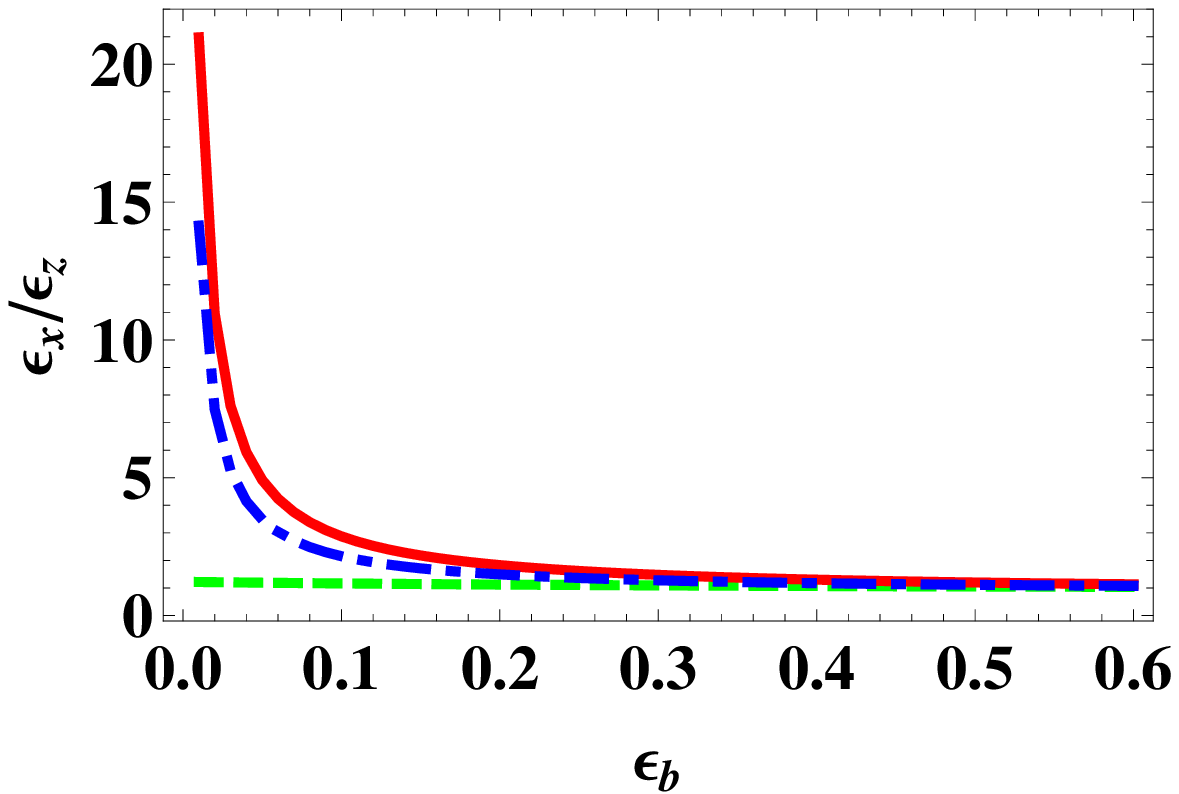}\hspace{10mm}
\includegraphics[width=2.5in]{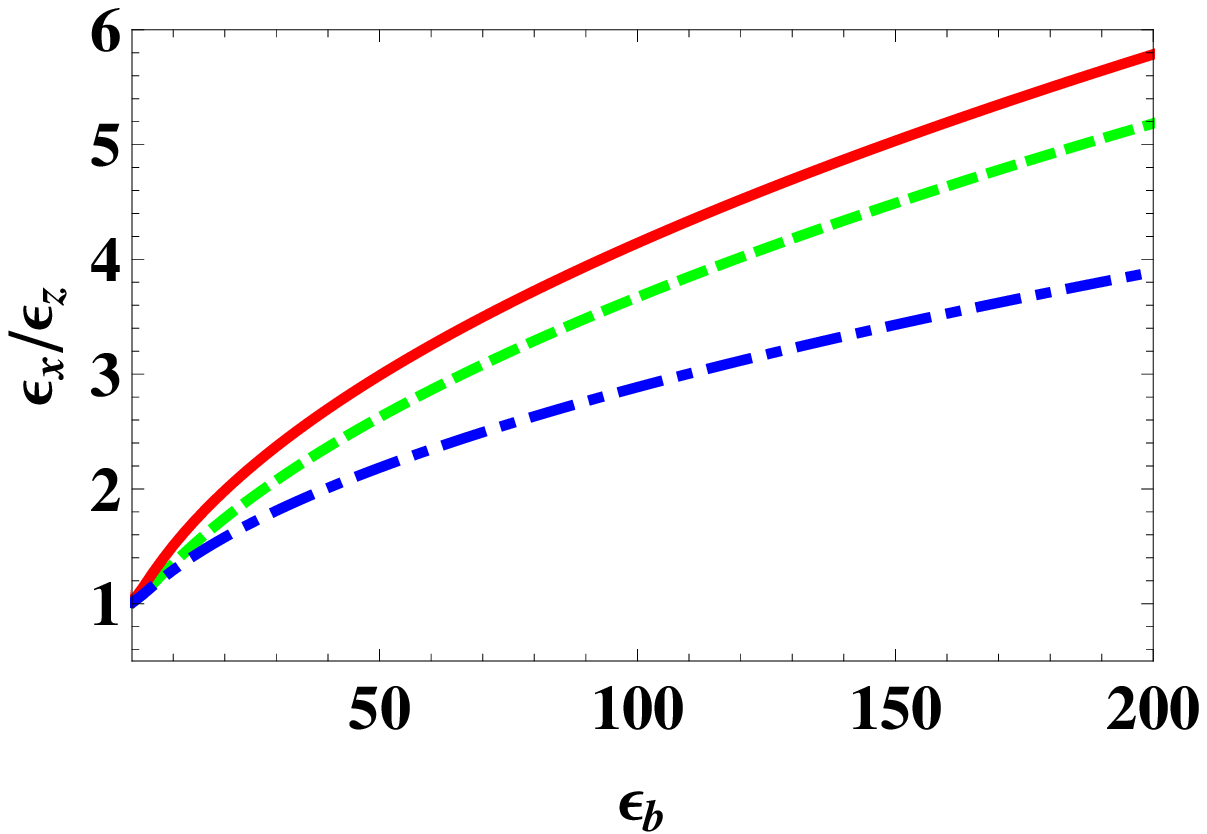}\\
 \caption{$\eps_x / \eps_y$, $\eps_y / \eps_z$ and $\eps_x / \eps_z$  plotted versus  $\eps_b \in \le 0.01, 0.6 \ri$ (left) and $\le  0.6, 200 \ri$ (right) for volume fractions $f_a = 0.7$ (green, dashed curves), 0.4 (red, solid curves) and 0.1 (blue, broken dashed curves). The component $a$ discs are  parallel to the $xy$ coordinate plane whereas the component $b$ discs are  parallel to the $xz$ coordinate plane.
  \l{fig6}}
\end{figure}

\newpage

\begin{figure}[!ht]
\centering
\includegraphics[width=2.5in]{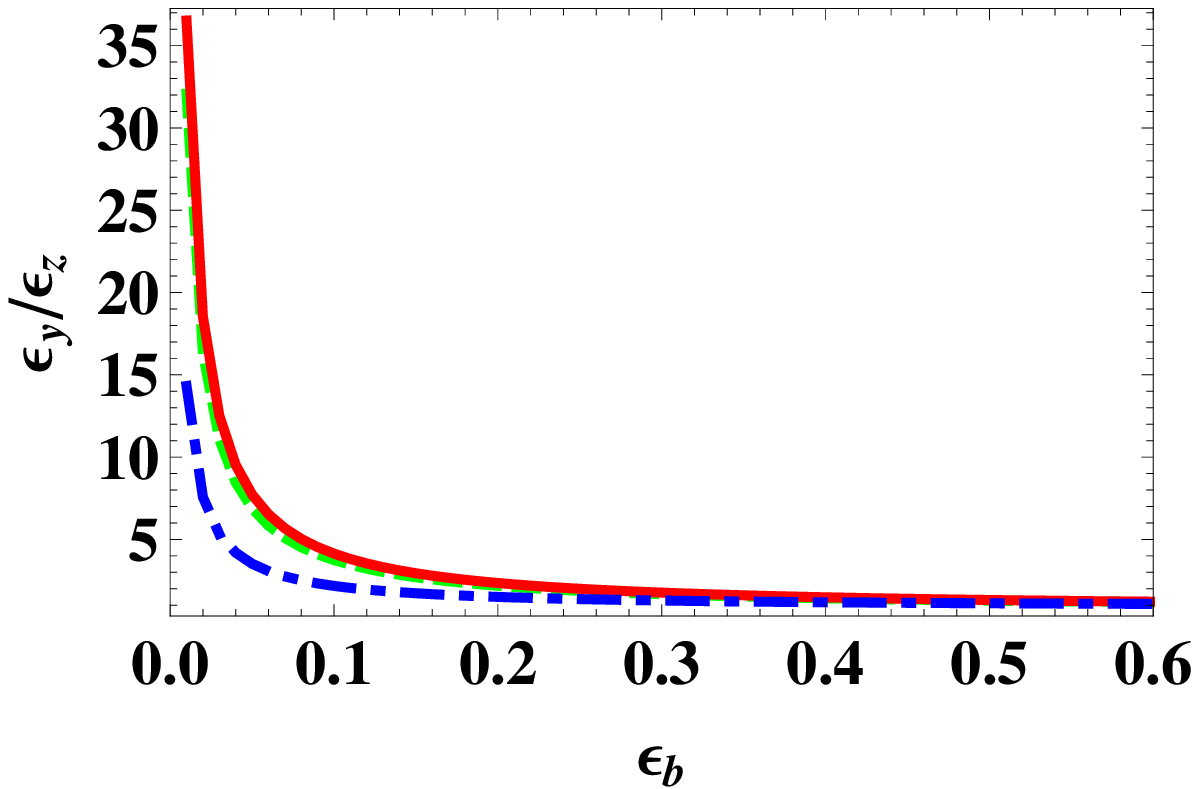}\hspace{10mm}
\includegraphics[width=2.5in]{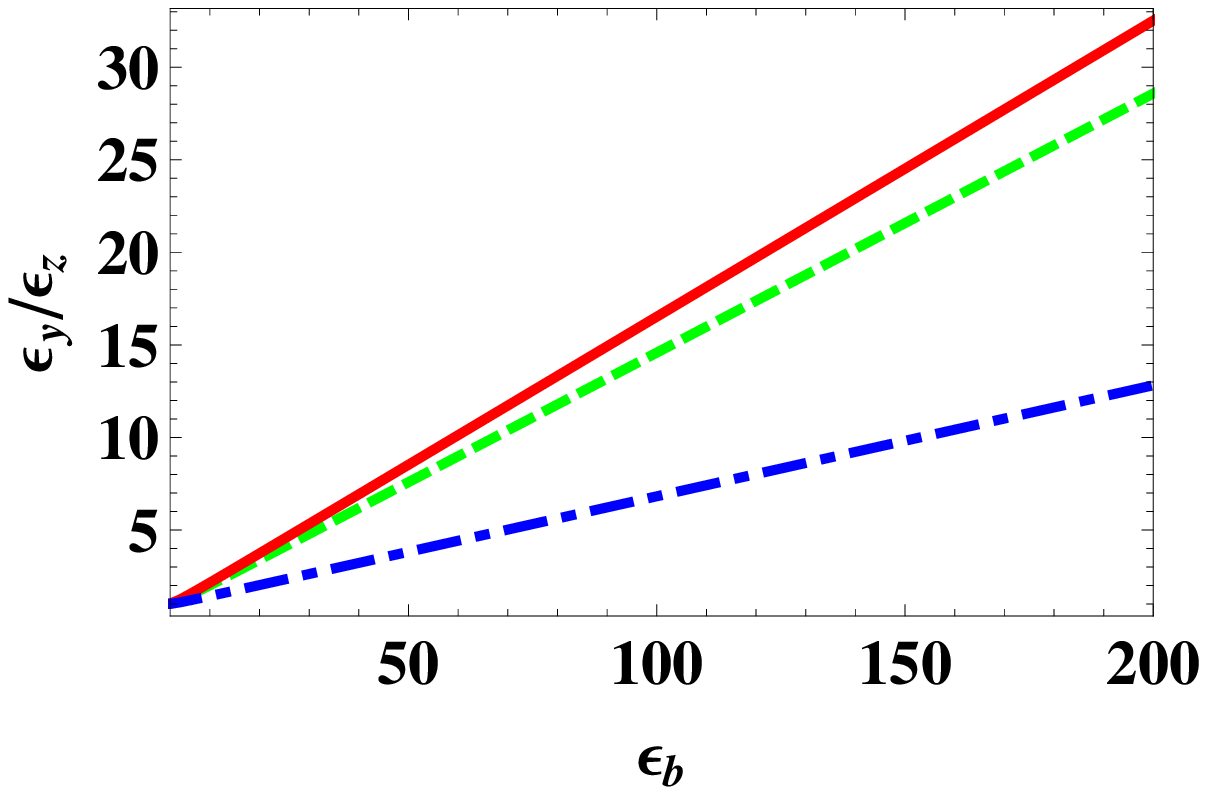}
 \caption{$\eps_y / \eps_z$ plotted versus  $\eps_b \in \le 0.01, 0.6 \ri$ (left) and $\le  0.6, 200 \ri$ (right) for volume fractions $f_a = 0.7$ (green, dashed curves), 0.4 (red, solid curves) and 0.1 (blue, broken dashed curves). The component  $a$ and $b$ discs are both  parallel to the $xy$ coordinate plane.
  \l{fig7}}
\end{figure}

\end{document}